\documentclass[letterpaper]{article} 
\usepackage[]{aaai2026}  
\usepackage{times}  
\usepackage{helvet}  
\usepackage{courier}  
\usepackage[hyphens]{url}  
\usepackage{graphicx} 
\usepackage{natbib}  
\usepackage{caption} 
\usepackage{algorithm}
\usepackage{algorithmic}
\usepackage{graphicx}
\usepackage{booktabs}
\usepackage{longtable}
\usepackage{ltcaption}

\usepackage{amsmath,amssymb,amsthm}
\newcommand{\CC}{\mathcal{C}}

\newcommand{\PP}{\mathcal{P}}

\newcommand{\Acal}{\mathcal{A}}

\newcommand{\Tcal}{\mathcal{T}}
\newcommand{\1}{\mathbb{1}}

\usepackage{enumitem}
\usepackage{tabularx}

\usepackage{xcolor}
\newcommand{\answerYes}[1]{\textcolor{blue}{#1}} 
\newcommand{\answerNo}[1]{\textcolor{teal}{#1}} 
\newcommand{\answerNA}[1]{\textcolor{gray}{#1}} 
 
\usepackage{newfloat}
\usepackage{listings}
\DeclareCaptionStyle{ruled}{labelfont=normalfont,labelsep=colon,strut=off} 
\floatstyle{ruled}
\newfloat{listing}{tb}{lst}{}
\floatname{listing}{Listing}
\title{A conceptual framework for ideology in online discourse beyond the left and right}

\author{
Kenneth Joseph\textsuperscript{\rm 1},
Kim Williams\textsuperscript{\rm 2},
David Lazer\textsuperscript{\rm 3}
}
\affiliations{
\textsuperscript{\rm 1}University at Buffalo\\
\textsuperscript{\rm 2}Portland State University\\
\textsuperscript{\rm 3}Northeastern University\\
kjoseph@buffalo.edu, kmw3@pdx.edu, d.lazer@northeastern.edu
}

\usepackage{bibentry}

\begin{document}

\maketitle

\begin{abstract}

Computational social science (CSS) has largely operationalized ideology along a single left/right partisan axis when studying online discourse. This approach obscures how people interpret and engage with more specific ideological formations related to race, climate, gender, and other domains. We introduce a framework that instead conceptualizes ideology as a multi-level socio-cognitive concept network and then explain how this \emph{conceptual model} of ideology can be linked to the study of online discourse. In doing so, our framework clarifies  how ideology manifests in discourse alongside related social processes such as framing, and provides an argument for better understanding of when and why we might study multiple concepts, such as values and beliefs, together in one analysis. More broadly, it bridges  methods used to study online discourse with ideology theory, enabling richer analyses of social discourse that benefit both fields. Although primarily theoretical, we provide a concrete operationalization of the framework on a sample of data, allowing us to show one example of how it can be used on real data.
\end{abstract}

\section{Introduction}

\begin{quote}
\small
    \emph{I'm sick of seeing tweets putting all us white people into one box. It isn't all of "us". Can't fix racism with more racism.}
\end{quote}

\begin{quote}
\small
    \emph{Are you part of the solution or the problem? I believe Black Lives Matter. I believe we can collectively make change.}
\end{quote}

The murder of George Floyd in 2020 prompted an eruption of social media posts about race. Many of these resembled the tweets above.\footnote{Modified to protect anonymity} Data from this period has been widely analyzed to understand discursive patterns across the left/right partisan divide \cite{shugarsPandemicsProtestsPublics2021}, as well as how discussion around specific issues evolved \cite{gallagherDivergentDiscourseProtests2018,freelonHashtagsFergusonBlacklivesmatter2016}. 
However, both analyses---the one that reduces individuals to a position on a spectrum, and the one that studies attitudes in isolation---obscure the fact that these two posts reflect a \emph{socially shared} and \emph{structured} way of understanding race  \emph{specifically}. That is, each example alludes to one of many views embedded in a temporally stable and socially shared \emph{system of thought} surrounding racial identity, the nature and/or existence of racism, and appropriate responses to inequality.


These specific, structured, temporally stable, and socially shared systems of thought are known as \emph{ideologies} \cite{freeden2021ideology,gerringIdeologyDefinitionalAnalysis1997}. Ideologies shape how we interpret, rationalize, and act on social issues such as race, climate change \cite{leeClimateChangeBelief2024}, and gender \cite{davisGenderIdeologyComponents2009}. 
Discourse, especially online, plays a central role in this shaping process, in that ideologies are often constructed, learned, and contested through it \cite{greenRhetoricalWhatGoes2024}. This suggests a natural role for studies of online discourse in the study of ideology: if ideologies shape interpretation and behavior \cite{jostPoliticalIdeologyIts2009}, and (online) discourse is a primary site where ideologies are constructed and learned \cite{vandijkIdeologyMultidisciplinaryApproach1998}, then the methods that help us study discourse should be a critical tool in the study of ideology.

Unsurprisingly, then, CSS has long studied the problem of identifying left/right ideological leanings in text  \cite{laverExtractingPolicyPositions2003}. More recent work has extended this paradigm for example by introducing multi-dimensional ideological spaces \cite{rottgerPoliticalCompassSpinning2024} 
and alternative axes (e.g., democratic versus authoritarian) \cite{piedrahitaDemocraticAuthoritarianProbing2025}. But even these alternative dimensional representations are blunt instruments \cite{levineMappingIdeologiesNetworks2024}. Individuals hold (their own interpretations of) many different ideologies which emerge in discourse in ways that  dimensional models struggle to capture \cite{maynardConvergenceDivergenceStudy2018}. For example, while the tweets above can be interpreted within a left/right framing of ideology, they are better understood in the context of ideologies \emph{centered on race}. The first tweet reflects elements of a \emph{colorblind} racial ideology \cite{bonilla-silvaRacismRacistsColorblind2006}, which broadly holds that the United States is a ``post-racial'' society. The second reflects an \emph{intersectional} ideology, which understands identity as multifaceted, that structural conditions are the primary cause of inequality, and that the margins of society must be centered when advocating for social change \cite{collinsIntersectionality2020}. 

As with all ideologies, however, the precise scope and boundaries of colorblindness and intersectionality are contested and fluid \cite{hamiltonAssessingExtendingColorblind2022,perezRacismHatredRacist2017}. Moreover, in the context of race \cite{mueller_racial_2020} 
and other social issues such as climate change \cite{leeClimateChangeBelief2024}, new ideologies can emerge, as can new discursive patterns that hold different ideological meanings. \textbf{Studying ideology beyond the left/right paradigm therefore faces a three-pronged challenge: (1) identifying which ideologies \emph{exist}, (2) identifying the \emph{contents} of a particular ideology (in a given socio-temporal context), and (3) determining which discursive patterns represent \emph{adherence} to which (parts of) which socially shared ideologies.} Complicating this task further are ongoing debates about what ideology even \emph{is} \cite{gerringIdeologyDefinitionalAnalysis1997,knightTransformationsConceptIdeology2006} and how ideologies are expressed in discourse alongside other processes like framing \cite{oliverWhatGoodIdea2000}.  

Identifying ideologies, their contents, and their discursive patterns is thus an important but  challenging task. Of course, there is no need to start from scratch. Social theory provides foundations for modeling ideology \emph{conceptually}---that is, in ways that go beyond spatial representations but that provide some abstraction from raw discourse (e.g. surface forms) \cite{maynardConvergenceDivergenceStudy2018}. Political scientists and social psychologists have long operationalized this theory through the lens of belief systems \cite{converseNatureBeliefSystems1964}, with  recent mathematical models of these systems suggesting new ways of quantifying ideology through a network perspective \cite{dalegeNetworksBeliefsIntegrative2024,warnckeResINNewMethod2026,vlasceanuNetworkApproachInvestigate2024}. However, such models have to date been developed and applied almost exclusively with survey data (or in simulation); work focused on how to infer ideological patterns \emph{from discourse} is limited \cite{homer-dixonComplexSystemsApproach2013} and challenging to expand to new settings  \cite{maynardMapFieldIdeological2013}. In contrast, the CSS community is capable of replicable and scalable analyses of ideologically-motivated discourse, but does not, at present, have the requisite theoretical clarity to specify how these tasks can help us to study ideology conceptually.

The goal of this paper is to build this link between the rich empirical landscape of online discourse and the powerful but empirically under-studied \emph{in discourse} conceptual models of ideology. To do so, we first outline a framework that 1) provides a succinct conceptual model that defines ideology as a \emph{multi-level network model}, 2) explains how ideology relates to other constructs that are widely studied in the CSS literature (e.g. framing), and 3) guides understanding of how ideology and other processes combine to shape and be shaped by discourse. In doing so, we build on work in discourse studies \cite{vandijkIdeologyMultidisciplinaryApproach1998}, ideology studies \cite{maynardConvergenceDivergenceStudy2018}, political science \cite{gerringIdeologyDefinitionalAnalysis1997,converseNatureBeliefSystems1964,greenRhetoricalWhatGoes2024}, psychology \cite{jostPoliticalIdeologyIts2009,dalegeNetworksBeliefsIntegrative2024} and sociology \cite{oliverWhatGoodIdea2000}. After outlining our framework, we present a brief case study that operationalizes it to study racial ideology on Twitter, showing one way in which it can be put to use on real world data.


\section{A Framework for Studying Ideology Conceptually in Discourse}

Almost all definitions of ideology\footnote{Table~\ref{tab:ideology_defs} in the appendix provides a few representative definitions, a more extensive set is constructed by \citet{gerringIdeologyDefinitionalAnalysis1997}.} invoke what  \citet{gerringIdeologyDefinitionalAnalysis1997} calls \emph{coherence}, the idea that an ideology is a \emph{collection of concepts that a group of people believe ``fit together.''} Scholars also agree that ideologies are not only interpretive frameworks; they are \emph{used, consciously or subconsciously,} to understand, motivate, and interpret behaviors and outcomes  \cite{freeden2021ideology,vandijkIdeologyMultidisciplinaryApproach1998}.  There is thus wide agreement that ideologies are, at minimum, shared and coherent worldviews that explain why ``it is how it is'' and ``what should be done about it'' \cite{oliverWhatGoodIdea2000}.  

However, this common definition leaves a gap between definition and operationalization. It does not, for example, specify what concepts we should be interested in or when some combination of concepts constitutes a coherent ideology. Answers to these questions vary widely.  For example, most agree that ideologies minimally include coherent systems of \emph{beliefs} \cite{converseNatureBeliefSystems1964,vandijkIdeologyMultidisciplinaryApproach1998}. However, social identities are treated by some as components of ideology \cite{boutylineBeliefNetworkAnalysis2017}, and by others as a distinct analytical dimension \cite{orrAffectivePolarizationDriven2023}. And even those who agree that ideologies consist only of beliefs often disagree on what exactly a belief is \cite{kalmoeUsesAbusesIdeology2020}. In an attempt to bring clarity to this operational challenge, other scholars have more recently begun to argue for incorporating multiple concept types (e.g. beliefs and values) into definitions of what an ideology is \cite{fishmanChangeWeCan2022,vlasceanuNetworkApproachInvestigate2024}. We build on this work here, linking a model of ideologies that incorporates multiple concept types in a multi-layer network with the study of online discourse.

We emphasize, however, that our framework does not aim to provide a definitive answer to which concepts should be modeled, or several other important questions about the nature of ideology. Instead, what it does do effectively is to \emph{provide a starting point for studying ideology conceptually in discourse that is consistent with an interdisciplinary literature} and that \emph{has strong links to existing CSS literature}. We do so by addressing four key questions:
\begin{enumerate}[nolistsep]
    \item Which concepts might an ideology consist of?
    \item If ideologies are networks, how should we conceive of relationships between concepts?
    \item Which criteria should be used to determine whether a given network of concepts forms a coherent ideology?
    \item When does(n't) an actor's discourse tell us they align with a particular ideology?
\end{enumerate}

\subsection{Which concepts might be part of an ideology?}
\begin{table*}[t]
\footnotesize
\setlength{\tabcolsep}{2pt}
\renewcommand{\arraystretch}{1.}
\centering
\begin{tabularx}{\textwidth}{
    @{}
    >{\raggedright\arraybackslash}p{1.9cm}
    >{\raggedright\arraybackslash}p{9.cm}
    >{\raggedright\arraybackslash}X
    @{}
}
\toprule
\textbf{Concept} & \textbf{Definition} & \textbf{Example}\\
\midrule
\multicolumn{3}{@{}l}{\textbf{Ideological Concepts}}\\
\midrule

Value &
[A]bstract ideals that guide conduct
\cite{maioIdeologiesValuesAttitudes2006} and ``that [glue together] more specific attitudes and beliefs''
\cite[][pg.211]{converseNatureBeliefSystems1964}
& Collective liberation; Freedom of the markets \\

Attitude &
``Tendencies to evaluate an object positively or negatively''
(pg. 283) that are a function of beliefs and emotions
\cite{maioIdeologiesValuesAttitudes2006}
& Support for Affirmative Action; Opposition to new climate taxes \\

Belief &
A determination about the truth of a statement
\cite[][pg.445]{vlasceanuNetworkApproachInvestigate2024}
or proposition \cite{vandijkIdeologyMultidisciplinaryApproach1998}
& Police disproportionately target people of color; Human activity is the main cause of climate change \\

Rationalization &
[R]easons that justify the truth of a belief
\cite[][pg.7]{williamsIdentityDefiningBeliefsSocial2022}
& `Race (Climate)-neutral’ policies such as credit scores reproduce racial gaps because they don’t account for race (existing warming)  \\

\multicolumn{3}{@{}l}{\textbf{Separate Dimensions of Analysis}}\\
\midrule

Expressed Issue Stance &
Stated preferences
& ``I support police reform''; ``I believe in climate change'' \\

Social Identity &
An often-named social position, role, group membership, or category that can be used to understand groups of individuals
& Republican; Scientist \\

Frame &
The result of a framing process that ``select[s] some aspects of a perceived reality and make them more salient in a communicating text''
\cite{entmanFramingClarificationFractured1993}, cf.\ \cite{cardMediaFramesCorpus2015}
& Social justice-oriented language to justify police or climate reform \\

\bottomrule
\end{tabularx}
\caption{Concepts often discussed in the context of ideology, separated between those that are elements of ideology and those that are not. Examples are given in the context of policing and climate change.}
\label{tab:concept_defs}
\end{table*}

Table~\ref{tab:concept_defs} presents a set of concepts that are elements of ideologies, as well as three constructs frequently analyzed in online discourse that are not.  Each is presented alongside examples in the context of police reform and climate change. Table~\ref{tab:concept_defs} is not exhaustive in two respects. First, the definitions provided do not reflect the full range of how these concepts have been theorized. Some argue, for instance, that values and attitudes are specific forms of beliefs \cite{kalmoeUsesAbusesIdeology2020}. Second, it does not capture all constructs that have been associated with ideology; we enumerate several of these in the appendix.  As with other closely related models of socially shared belief systems, then, the ``levels and the characteristics of the different constructs are not exhaustive. Rather, they are provided as examples of dimensions on which these constructs may vary'' \cite{vlasceanuNetworkApproachInvestigate2024}. While thus not exhaustive, Table~\ref{tab:concept_defs} does capture the four concepts most often assumed to constitute ideology: values, attitudes, beliefs and their rationalizations. 

These concepts can be arranged into a partial hierarchy. At the bottom are values, which are widely understood to be fewer, more abstract, and more prescriptive than attitudes, beliefs, or rationalizations \cite{maioIdeologiesValuesAttitudes2006,oliverWhatGoodIdea2000,vandijkIdeologyMultidisciplinaryApproach1998}. At the top of the hierarchy lie rationalizations, which are understood to be the most concrete and that serve as justifications for beliefs  \cite{williamsMarketplaceRationalizations2022,vlasceanuNetworkApproachInvestigate2024}. In between lie beliefs and attitudes. Values often are understood as informing attitudes \cite{vishwanathImpactValuesIssue2025}, and we thus place attitudes directly above values. Beliefs sit above attitudes, although their relationship may be best understood as symbiotic rather than hierarchical \cite{fabrigar2005structure}. Our hierarchy thus represents structure in terms of which concepts might inform which other types of concepts. It is not meant to imply different degrees of stability, that one level is a prerequisite for another, or, in the case of attitudes and beliefs, at least, differences in prevalence.

Table~\ref{tab:concept_defs} also lists three concepts often examined in CSS that ideologies do \emph{not} consist of: social identity, expressed issue stance, and frames. Ideologies interact with identity-driven and framing processes to produce discourse, where stances on issues are expressed. Expressed issue stances are thus \emph{discursive behaviors}, not direct manifestations of ideological concepts. Interpretation of expressed stance in discourse must therefore account for mediating processes, a claim that is true even in structured contexts like surveys \cite{greenSpeechActPerspective2020,berinskyMeasuringPublicOpinion2017}. Frames are not components of ideologies because framing is the act of selecting and emphasizing particular aspects of a situation or issue, while ideology refers to the underlying system of meaning being communicated \cite{oliverWhatGoodIdea2000}. Not all CSS scholars see such a clean separation between frames and ideology, see e.g. \citet{sullivanThreeLevelsFraming2023}.
  But separating frames from ideology allows us to ask a richer set of questions like, what \emph{set of} ideologies are embedded in the frames that are typically induced by a communicative content \cite{vandijkIdeologyMultidisciplinaryApproach1998}?  Finally, social identities often exist in concert with specific ideologies or even specific ideological concepts \cite[e.g. sacred value theory, ][]{tetlockThinkingUnthinkableSacred2003}
. But identities are who we (think or subconsciously feel we) are, and ideologies are what we (think we should or are subconsciously motivated to) believe.  Preserving this distinction is important because identities are not inherently ideological \cite{masonIdeologuesIssuesPolarizing2018}, and because doing so enables researchers to examine how identities and ideologies jointly shape attitudes and behavior across contexts \cite{kalmoeUsesAbusesIdeology2020}.

\subsection{What about relationships between concepts?}


Ideologies, however, do not only consist of concepts, they also prescribe how these concepts are \emph{related}. Empirical research on ideology largely focuses on socially-shared cognitive associations, where a view on one concept is expected to bring to mind a view on another \cite{converseNatureBeliefSystems1964}. Most work identifies these links between concepts using population-level pairwise statistics (e.g. the correlation between attitude responses in surveys) \cite{converseNatureBeliefSystems1964,fishmanChangeWeCan2022}. These relationships are, on their surface, tautological---if ideologies are socially shared, then concepts within the same ideology should be correlated within populations. But some concepts within ideologies are more strongly connected, both in the way they are internalized \cite{fishmanChangeWeCan2022} and how they are expressed in discourse  \cite{wanPoliticalPunditsMaintenance2026}.  Ideology researchers therefore do expect some of these relationships to be stronger or more important than others, and further anticipate that this structure is important in understanding individuals' ideological sophistication \cite{converseNatureBeliefSystems1964}. 

Again, however, different theories present different assumptions about where links may (not) exist between concepts. For example, some models of ideology restrict relationships to be hierarchical, linking increasingly specific beliefs down a chain of beliefs \cite{peffleyHierarchicalModelAttitude1985}. This idea has more recently been extended to emphasize hierarchical links between concept types, including those referenced above between values and attitudes, and between beliefs and rationalizations. These recent works thus suggest that ideologies ``are perhaps best understood to have both hierarchical and more horizontal structure'' \cite{homer-dixonComplexSystemsApproach2013}.  More concretely, this relational structure means that \textbf{ideologies can be formalized as a multi-level network structure, where links exist both within concept types and, hierarchically, between successive types.}



\subsection{What counts as a coherent ideology?}\label{sec:coherence}

Our framework defines four types of coherence an ideology should satisfy: \emph{subject matter} coherence, \emph{temporal} coherence, \emph{social} coherence, and \emph{conceptual} coherence.\footnote{Our treatment of coherence differs from \citet{gerringIdeologyDefinitionalAnalysis1997}, while remaining consistent with his definition of ideology. Whereas \citet{gerringIdeologyDefinitionalAnalysis1997} uses coherence primarily in the conceptual sense, he also recognizes temporal stability, subject matter coherence, and shared-ness as important properties. We broaden coherence to encompass these properties as assumptions that can constrain the learning of ideological structure from discourse.}

\paragraph{Subject matter coherence} Ideologies exist around particular domains of social life where group interests conflict \cite{vandijkIdeologyMultidisciplinaryApproach1998}. CSS researchers should therefore be explicit about the domain they are studying and use it as a scope condition for analysis. For example, those studying political ideologies should state what they take to be ``political'' and what that choice assumes \cite{laneWhatPoliticalExpression2022}. The requirement, however, is that an ideology be coherent \emph{within} the domain of study, not that it be confined to it. Intersectionality, for instance, spans many subject areas, but we study it here only as it bears on race: the concepts, relations, and adherents we recover are those expressed within that domain, and we make no claim about its structure elsewhere.

\paragraph{Temporal coherence} Ideologies change over time, but retain a core set of concepts and relations that persist across those changes \cite{vandijkIdeologyMultidisciplinaryApproach1998}. CSS researchers are familiar with this kind of assumption from work modelling word meanings as drift around a stable core \cite{hamiltonDiachronicWordEmbeddings2016}. Ideologies complicate this assumption, however, because concepts at different levels of abstraction surface differently in online discourse. More abstract concepts, like values, exert a pervasive but diffuse influence on discourse, supporting frequent but noisy inferences about which actors align with those values \cite{borensteinInvestigatingHumanValues2025}. Less abstract concepts, like rationalizations, are expressed rarely but are easier to identify when they appear, yielding sparse but precise measurements of concept alignment. There is thus a tension between how often we can measure a concept, how well we can measure it, and how much of the ideology's internal structure that measurement preserves. \citet{berinskyMeasuringPublicOpinion2017} confronts a version of this in survey research, arguing for meso-level constructs general enough to measure reliably but specific enough to capture the nuances and dynamics of survey response. Two things differ when studying online discourse. First, the tradeoff runs in the opposite direction: in surveys, more general scales aggregate over question-level noise and so measure more consistently, whereas in discourse generality \emph{costs} reliability, for the reasons just given. Second, while surveys are measured at fixed points, a concept is observed in discourse only when someone happens to express it, so apparent temporal dynamics are driven both by how often a concept is expressed at all and by exogenous factors shaping when people post. Measures of temporal stability are thus entangled with external factors in a way they are not in surveys.

Rather than settling on a single meso-level construct, then, we argue for modelling concepts at several levels of abstraction within a single ideology, and letting them inform each other. Frequent signals from abstract concepts can help guide how (un)likely it is that we will obtain a sparse observation of a belief or rationalization, while the rare concepts, once found, can help to anchor and inform the dynamics of more abstract measurements. To the extent that exogenous factors impact all concepts, differential stability then becomes evidence rather than noise: levels moving together points to a change in those conditions, whereas divergence between them points to a change in the ideology itself.

\paragraph{\emph{Partial} social coherence} Ideologies must be understood as socially shared \cite{jostPoliticalIdeologyIts2009}, because they are collectively constructed through discourse \cite{leadermaynardDangerousSpeechDangerous2016} and strategically manipulated in discourse by political actors \cite{greenRhetoricalWhatGoes2024,wanPoliticalPunditsMaintenance2026}. What most shapes behavior, moreover, is what people understand their ideological positions \emph{should be}---i.e. ideological \emph{norms} \cite{groenendykHowNormsShape2023}. But that sharing is partial in two distinct senses. At the group level, ideologies are not universally shared because they operate in spaces where interests and/or power are contested \cite{vandijkIdeologyMultidisciplinaryApproach1998}. This is what distinguishes ideology from culture and commonsense, which are shared meanings that \emph{allow interaction across groups}, whereas ideologies are leveraged by groups to compete for and/or rationalize power \cite{vandijkIdeologyMultidisciplinaryApproach1998}. A belief shared by everyone is thus not ideological \cite{vlasceanuNetworkApproachInvestigate2024}. At the individual level, an ideology may be coherent for the group while any individual expression of it is a sample, filtered through what that person received, accepted, and finds appropriate to say in context \cite{converseNatureBeliefSystems1964,zaller1992nature}, or what they believe is strategically effective \cite{wanPoliticalPunditsMaintenance2026}. Social coherence must thus be informed by a clear theoretical understanding of the link between individual ideological expression and the social nature of ideology, a point we expand on below.

\paragraph{Conceptual coherence} Concepts within an ideology should be \emph{structurally} and \emph{logically} coherent. By structurally coherent, we mean that concepts should have more and/or stronger relations to other concepts within the ideology than to concepts outside of it. Prior work on ideology has examined its network structure from multiple angles: identifying central concepts within ideologies \cite{boutylineBeliefNetworkAnalysis2017,warnckeWhatExplainsCountryLevel2025,fishmanChangeWeCan2022}, testing whether subpopulations exhibit certain levels of conceptual coherence,\footnote{This has been a central question in American political science for sixty years \cite{converseNatureBeliefSystems1964}, focusing on whether \cite[and how many][]{kalmoeUsesAbusesIdeology2020} Americans hold coherent liberal or conservative ideologies.} and qualitatively identifying ideological structure from discourse where all text is assumed to align with a single ideology \cite{homer-dixonComplexSystemsApproach2013,levineMappingIdeologiesNetworks2024}. However, virtually no work in this space has explored the task of being given a text corpus and then \emph{discovering whether zero or more coherent ideologies are expressed within it}. By logical coherence, we simply mean, for example, that an ideology should not represent both support for and opposition to affirmative action.   Note that this is different from assuming that ideologies should be logical in the sense that they reflect the world as it truly is. Many ideologies plainly do not. 

\subsection{When does an actor adhere to an ideology?}

Ideology manifests discursively in bits and pieces, through context-specific expressions from many actors that are shaped by how those actors identify themselves, frame their message, and have interpreted the ideology in context \cite{vandijkIdeologyMultidisciplinaryApproach1998}. This leads to two intertwined complications in determining when actors align with an ideology given their discourse: 1) inferring the expression of ideological concepts and/or their relationships from discourse, and 2) using those inferences to then infer coherent ideologies and their adherents.

\paragraph{(How) can we infer ideological concept expression in discourse?}
Table~\ref{tab:concept_extraction} in the appendix provides examples of specific recent papers in the CSS literature that focus on the inference of the concepts we model from text. For example, \citet{introneMeasuringBeliefDynamics2023} and \citet{burnhamStanceDetectionPractical2025} present methods for open-domain inference of beliefs and attitudes, respectively. One can also, as we do below, pre-define a set of concepts and then leverage more general means for classification \cite{haltermanCodebookLLMsEvaluating2026}. Combining these approaches to concept extraction with a model of coherent ideological structure provides interesting opportunities to extend and improve extraction. One example of this, mentioned above, is that the joint inference of more abstract concepts, like values, and more concrete but sparsely observed signals, like rationalizations, may complement each other. Another, which we discuss further below, is allowing us new ways to assess potential inaccuracies in concept extraction by comparing predictions to a theoretically informed model. 

\paragraph{(How) do (we know when) actors ``have'' an ideology?}
Given a set of actors and their associations with specific ideological concepts, identities, and frames, when do we say an individual ``has'' an ideology, when we may not even know which ideologies exist? This is a form of the well-known dual problem in sociology \cite{breigerDualityPersonsGroups1974}, and emphasizes that in order to draw inferences about the link between individual and ideology, we must either leverage existing understanding of our domain and assumptions about coherence, or accept that we can capture only fragments of ideological thought in socio-temporally situated data. 

As insinuated above, one critical coherence assumption revolves around partial social coherence, and specifically how socially shared ideologies emerge as a product of individual decisions and discourse. Theoretical work on the link between individual and social ideological structures largely concerns how and to what extent ``ordinary'' people internalize and then reproduce shared ideological structures. Most notably, Zaller's (1992) RAS model argues that individuals \emph{R}eceive political messages through their social and informational environments, \emph{A}ccept or reject them depending in part on their existing predispositions, and \emph{S}ample from the resulting considerations when responding to surveys. In this account, survey responses are therefore constructed from accessible considerations rather than retrieved from stable, internally coherent ideological structures. 
\citet{greenSpeechActPerspective2020} extend the RAS model by situating survey responses within a \emph{speech act perspective} (SAP), where actors not only sample when responding but \emph{integrate} that sampling process into the social context constructed from the conversational exchange between researcher and respondent. 
Though conversations within surveys and online discourse differ substantially, Green et al.'s (2020) work offers a theoretical link between the RAS model and the context-specific sampling and expression of ideological concepts in a discursive setting.  

The SAP perspective also links to work on how ideologues who aim to shape and define ideologies express them in online discourse \cite{greenRhetoricalWhatGoes2024,wanPoliticalPunditsMaintenance2026}. These are the actors we study in our case study below as well. As the ones defining ideological boundaries, ideologues have a more concrete understanding of ideologies and how they want to shape them. Such efforts, however, are often narrowly scoped to ideological substructures that they have incentives to focus on \cite{wanPoliticalPunditsMaintenance2026}, and thus must also be studied collectively. It is worth noting as well that this empirical work on online ideologues is relevant to the settings where our proposed framework is useful; however, this existing work focuses on language itself as the unit of analysis. In contrast, we emphasize the utility of abstracting analysis to the conceptual level.  Taken together, RAS and SAP imply that the link between individual discourse and socially shared structure can only be recovered across many expressions and many actors, an assumption encoded in our assumed forms of coherence.


\section{An Example Operationalization}

As with related models, our framework is designed to accommodate varied study contexts and empirical approaches, leaving some decisions—such as which concepts to operationalize—open to the researcher. Other empirical questions, such as how best to identify ideological concept expression in text, are beyond the scope of this paper. We nonetheless provide an example operationalization to illustrate how the framework can be applied. 

Specifically, we build a straightforward approach, informed by our framework, to study ideological expression in a set of 62,045 tweets about race or race-related content sent between 2012 and 2022 by 607 Twitter users.  These users are a sample of a larger set of 33,811 users whose tweets about race and race-related content were retweeted at least once by a panel of 1.6 million Twitter handles matched to eligible voter data in the United States \cite{shugarsPandemicsProtestsPublics2021,hughes2021TwitterComparison}. Race-related tweets are identified using a high-precision, high-recall keyword-based method from \citet{shugarsPandemicsProtestsPublics2021}. Additional details about the dataset can be found in the appendix. Our goal is \emph{not} to provide novel or robust empirical findings. Doing so would require a more dedicated empirical analysis that is not the focus of this paper. Rather, we aim to demonstrate one way in which our framework might be operationalized and applied to real-world data. Code and anonymized data necessary to replicate the work in this section are provided at \url{https://github.com/kennyjoseph/conceptual-ideology}.

\subsection{Methods}

Here, we present an overview of how we infer ideological concepts expressed in tweets, how we use those inferences to infer coherent ideologies, and the evaluation metrics we use to assess model fit overall and across different parameterizations.  Additional details  can be found in the appendix.

\paragraph{Concept extraction from discourse} We adopt a straightforward approach to concept extraction, using zero-shot labeling with a pre-defined codebook. We first construct a codebook of concepts that align with one of three ideologies: intersectionality (IS), color-blindness (CB), and white supremacy (WS). 
Ideological concepts are identified through a combination of our prior grounded theory work \cite{williamsBlackLivesMatter2024} and refinement of those findings using a large-language model (Claude Fable 5.0) for the purposes of the zero-shot labeling performed here. The codebook contains 13 values (4 CB, 3 WS, 4 IS, 2 shared), 28 attitudes (4, 7, 10, 7 shared), 35 beliefs (5, 11, 11, 8 shared), and 36 rationalizations (10, 11, 10, 5 shared). After constructing this codebook, we then use zero-shot labeling, passing a detailed prompt (including the codebook) to \texttt{GPT OSS 20B} and asking it to provide, per tweet in our dataset, the concepts expressed, as well as a confidence in $[0,1]$. The appendix provides the full codebook and prompt.

\paragraph{Leveraging Concept Inferences to Infer Ideological Alignment and Structure} Our modeling approach draws heavily on the structural topic model (STM) \cite{robertsStructuralTopicModels2014} and the sparse generative additive model of text (SAGE) \cite{eisensteinSparseAdditiveGenerative2011}, where instead of modeling documents as a bag-of-words, we model documents as a bag of ideological concepts, and incorporate an observation model by which concept co-occurrence informs inferred ideological network structure. 

More specifically, we develop a statistical model of the process by which a set of actors $\Acal$, with metadata $m_a$, produce discursive expressions $d$, each of which may express zero or more codebook concepts $u$ at a time $t$. We then construct a pseudo-observation model for $A^{(\delta)}_{a,uu'}$, the co-occurrence between two concepts $u$ and $u'$ for some actor $a$ over some time window $\delta$. An ideology $k$ is a pair $(\eta_k, W_k)$, where $\eta_k$ is a sparse vector of rates at which each concept $u$ is expressed in the corpus by adherents of this ideology over and above some base rate, and $W_k$ is a block-structured, symmetric network that models concept co-occurrence driven by $k$ (and not chance). The structure of $W_k$ is fixed such that relationships can only exist within concept types and between successive rungs of the multi-level network (values to attitudes, attitudes to beliefs, etc.).

Actors are optionally linked to ideologies through metadata and are modeled as potentially specializing in particular \emph{facets} of ideologies.  Mathematically, then, actors in the model have a soft membership $\theta_a$ over (ideology, facet) pairs: each ideology has $F_k$ facets $\eta_{k,f}$ that share the ideology's single network $W_k$; we vary $F_k$ across model runs and compare the resulting fits below. As noted, our model allows for actor metadata to inform \emph{which ideology} an actor is associated with (in the same way the STM allows topic distributions to vary via metadata):
\[
  \theta_a=\mathrm{softmax}\big(\underbrace{\beta_k^\top m_a}_{\text{ideology logit}}+\underbrace{\gamma_{a,kf}}_{\text{facet logit}}\big),
  \quad \gamma_{a,\cdot}\sim\mathcal{N}(0,\sigma_\gamma^2 I).
\]
We write $\vartheta_{a,k}=\sum_f\theta_{a,kf}$ for the ideology-level membership (facets of an ideology share its network), and model $\theta_{a,0}$ as a background coordinate, characterizing the share of the actor's speech that is unideological. 

For each tweet $d$ sent by actor $a$ at time $t$, the zero-shot model described above estimates a confidence score $x_{d,u}\in[0,1]$ that $d$ expresses concept $u$. We treat these as expected counts (without thresholding) in a Poisson model with rate $r_{a,u}$. Tweets are thus a sparse set of the actor's concepts, each conditioned independently based on a rate parameter $r_{a,u}$, defined as:
\[
  \log r_{a,u}\;=\;\underbrace{m_u}_{\text{background}}\;+\;\underbrace{\textstyle\sum_{k,f}\theta_{a,kf}\,\eta_{k,f,u}}_{\text{ideology and facet}}\;+\;\underbrace{\mu_a}_{\text{breadth}}.
\]
The rate consists of the base rate of a concept overall in the dataset ($m_u$, fixed at the corpus rate), the rate defined by an ideology and the facet within it, and a scalar that accounts for actors with different rates of activity ($\mu_a$).

We then model co-occurrence of concepts over a pre-defined timescale $\delta$: for each actor $a$ and pair of concepts $(u,u')$, we define a kernel-weighted pair count over the actor's units,
\[
  A^{(\delta)}_{a,uu'}\;=\;\sum_{d,d'}K_\delta\big(t(d)-t(d')\big)\,x_{d,u}\,x_{d',u'},
\]
Here, we fix $\delta$ to be a year, and thus co-occurrence statistics are computed per year of the data (for each actor) and then \emph{treated as observed data}. We then model the likelihood of a given pseudo-observation $A^{(\delta)}_{a,uu'}$ as:
\[
\begin{aligned}
A^{(\delta)}_{a,uu'} &\sim \text{Poisson}\Big(N^{(\delta)}_a\;\bar r_{a,u}\,\bar r_{a,u'}\;e^{\Omega_{a,uu'}}\Big),\\
\Omega_{a,uu'} &= W^{(0)}_{uu'}+\sum_k\vartheta_{a,k}W_{k,uu'},\\
N^{(\delta)}_a &= \sum_{d,d'}K_\delta\big(t(d)-t(d')\big),
\end{aligned}
\]
where $\bar r_{a,u}$ is the actor's rate for that concept, and $W^{(0)}$ is a shared association layer, estimated from all actors, that accounts for associations common across \emph{all} actors (and not only those within a particular ideology); i.e. cultural, rather than ideological, associations. The objective to be minimized thus draws on a likelihood function conditioned on the observed (from the extractor) data $x_{d,u}$ and the pseudo-observation $A^{(\delta)}_{a,uu'}$. The model is estimated using a block-coordinate descent approach that mixes closed-form, gradient-based, and variational updates.

\paragraph{Evaluation} We consider four ways to evaluate model fit. First, in order to test predictive capacity, we consider \emph{held out fit}: we withhold a randomly sampled (consistent across parameterizations we consider below) ten percent of each actor's tweets. We then compute the difference between the concepts expressed in these tweets and the model's predictions using the (Poisson) loss of the fitted rates on the observed data. Second, while the concepts we study are initially defined by their alignment with one of three ideologies, \emph{the model itself does not know the assignment from concept to ideology}. We therefore can assess the model based on how well its final distribution of $\eta_{k}$ aligns with the researcher-defined ideology from which the concept draws. Because we obtain a continuous value of $\eta_{k}$ for each concept and each ideology, we used a rank-based area-under-the-curve (AUC) measure and report the macro-average, $\overline{AUC}$. Third, we consider the \emph{share} (percentage) of actors most strongly aligned with each ideology. Cases where no actors are aligned with an ideology signal model degeneracy.

Finally, rather than incorporate structural conceptual coherence into the model, we instead treat it as an evaluation measure that can be used to help identify ideologies identified by the model that have more/stronger relations internally than externally. Mathematically, for ideology $k$ having concepts $S_k=\{u:\max_f\eta_{k,f,u}>0\}$, we define a \emph{structural coherence ratio} $C_k$ comparing the relational strength for pairs of concepts where \emph{both} concepts are in $S_k$ to the relational strength where \emph{exactly one} concept in the pair is in $S_k$. Formally, $C_k \;=\;
    \frac{\sum_{(u,u')\in\PP,\; u\in S_k,\; u'\in S_k} W_{k,uu'}}
         {\sum_{(u,u')\in\PP,\; u\in S_k \,\oplus\, u'\in S_k} W_{k,uu'}}$, where $\PP$ is the set of allowed concept pairs (those not excluded by the type structure or by logical coherence).  The measure $C_k$ has a one-to-one correspondence with the well-established measure of weighted conductance from the network science literature \citep{bollobas1998modern}. Like in the study of conductance, we can use a permutation test to characterize whether or not the ratio $C_k$ is larger than we would expect by chance for a similarly structured random set of edges in $\PP$.

\paragraph{Summary and Identifiability} The model used in our working example encodes several aspects of our framework. It defines ideologies using the structured multi-level network model of concepts we propose. \emph{Subject matter} coherence is operationalized by restricting our analysis to race-related content using the approach from \citet{shugarsPandemicsProtestsPublics2021}; \emph{temporal} coherence is operationalized by assuming that ideologies are static over the entire time window; \emph{partial social coherence} is operationalized by treating actors as being a mixture over ideologies and probabilistically expressing concepts associated with that ideology, and by having actor's ideology optionally informed by identity signals in actor metadata; and \emph{structural conceptual} coherence is encoded with an evaluation metric that allows us to determine how structurally consistent an inferred ideology is relative to a random baseline distribution.

As defined to this point, however, the model does not incorporate logical conceptual coherence, and cannot identify theoretically established ideologies because it does not incorporate pre-defined signals of specific ideologies. To address the former, we draw on the ResIN framework of \citet{warnckeResINNewMethod2026} and model \emph{response-items}. Most concretely, this means that we model attitude valences separately, with (for example) separate concepts for ``Supports Affirmative Action'' and ``Opposes Affirmative Action.'' We then fix $W_{k,uu'}=0$ between the two valences of an attitude, so that no ideology can link a position to its own negation. To address the latter, we fix, or \emph{pin}, actors to ideologies. Specifically, we identify three actors that prior work establishes as being definitively aligned with each ideology we model, and use those to guide identifiability. As we show below, fixing these three actors, with a closed codebook, can effectively address issues of identifiability. 

In summary, we use a combination of theory (informed by our framework) and domain expertise to build a methodology to infer ideologies and their adherents. Much like the topic modeling literature on which our modeling approach draws, we also use a combination of held-out data, theory, and domain expertise in evaluation \cite{chen2023we}. The decisions we make could, of course, be extended, for example by allowing ideologies to change over time, or by incorporating network ties between actors. We touch on additional limitations in our conclusion. Nonetheless, our operationalization presents one concrete way to develop, test, and leverage conceptual models of ideology through a careful and explicit combination of social theory and data-informed modeling. 

\subsection{Results}
 \begin{table}[t]
      \centering
      \footnotesize
      \setlength{\tabcolsep}{2.5pt}
      \renewcommand{\arraystretch}{1.08}
      \begin{tabular}{@{}lrrlr@{}}
      \toprule
      Parameterization & Held-out & $\overline{\mathrm{AUC}}$ &
      $C_k$ & Share (\%) \\
      \midrule

      \multicolumn{5}{@{}l}{\textit{Two fixed ideologies, CB + IS, Vary number of facets}}\\[-2pt]
      \textbf{1 Facet/ideology (1F)}  & 21,680 & \textbf{.878} &
        \textbf{CB}, \textbf{IS} & \textbf{36/64} \\
      2F  & 21,468 & .812 &
        \textbf{CB}, \textbf{IS} & \textbf{45/55} \\
      3F  & 21,512 & .728 &
        CB, IS & \textbf{93/7} \\
            \addlinespace[2pt]
      \multicolumn{5}{@{}l}{\textit{Three fixed ideologies (CB, WS, IS)}}\\[-2pt]
      1F + WS  & 21,729 & .769 &
        \textbf{CB}, \textbf{WS}, \textbf{IS} & 0/28/72 \\
      2F + WS  & 21,542 & .717 &
        \textbf{CB}, WS, \textbf{IS} & 86/0/14 \\
      \addlinespace[2pt]
      \multicolumn{5}{@{}l}{\textit{Two fixed ideologies + ``free'' ideology}}\\[-2pt]
      1F + free ideology & 21,688 & .869 &
        \textbf{CB}, \textbf{IS}, free & 38/62/0 \\
      2F + free & \textbf{21,463} & .696 &
        \textbf{CB}, \textbf{IS}, free & 96/4/0 \\

      \bottomrule
      \end{tabular}
  \caption{Results for our four evaluation metrics from seven sample model runs across three setups. Shares are listed in the same order as the components named in the $C_k$ column. Bold coherence marks $C_k$ with $p<.01$ against the matched permutation null.}
  \label{tab:ladder}
  \end{table}

\begin{figure}
    \centering
    \includegraphics[width=\linewidth]{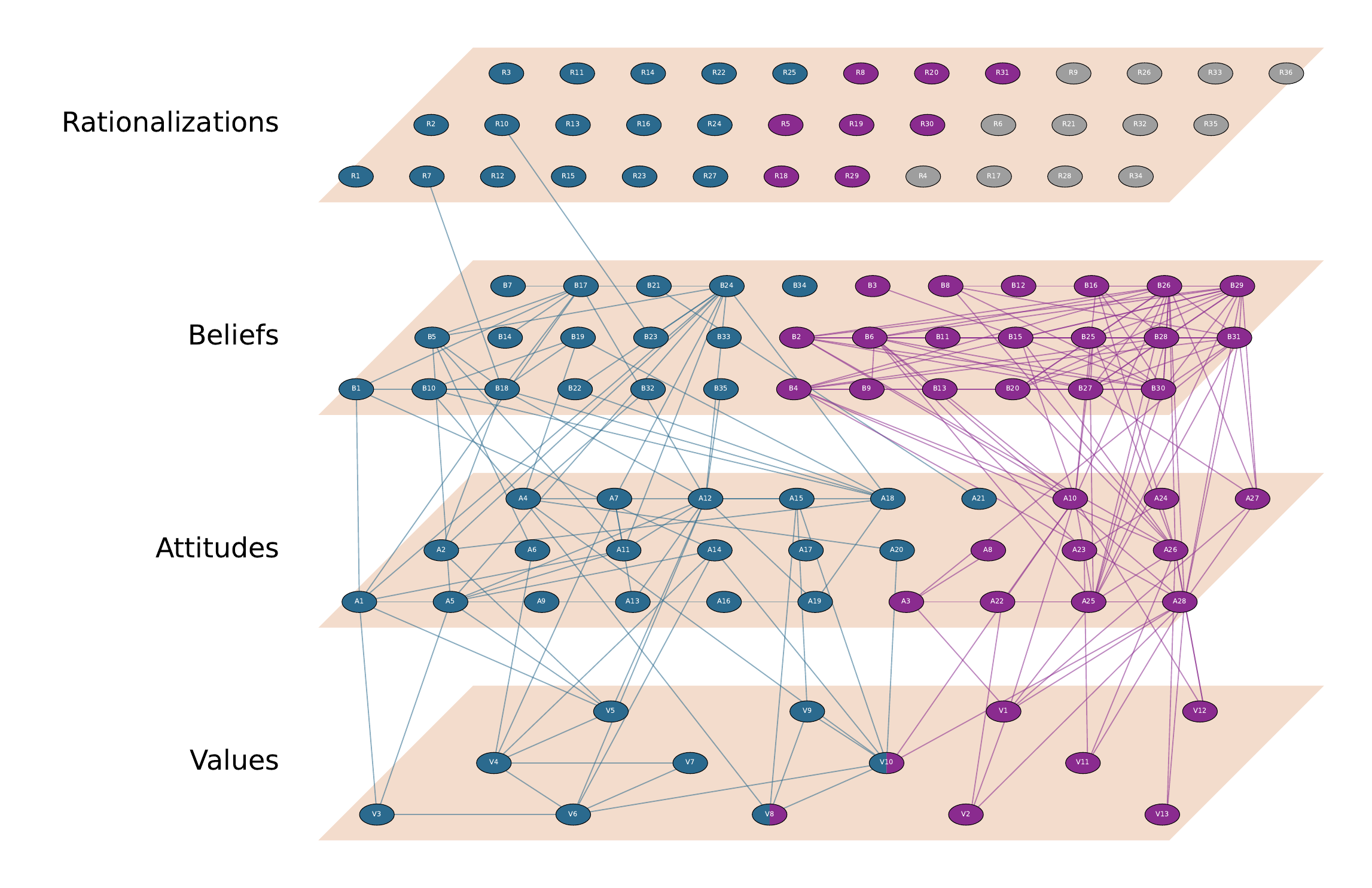}
    \caption{The two recovered ideologies over the four levels of ideological concepts. Blue is for color-blindness (58 concepts, 82 edges), purple is for intersectionality (44, 116). Split colors indicate nodes assigned to both ideologies, grey for the 12 concepts assigned to neither ideology.}
    \label{fig:model_out_layers}
\end{figure}

\begin{table}[t]
\centering
\footnotesize
\begin{tabular}{>{\raggedright\arraybackslash}p{6cm}lr}
\toprule
Concept & Type & $\eta$ \\
\midrule
\multicolumn{3}{@{}l}{\footnotesize\itshape Colorblindness} \\
Positive evaluation of a shared national identity over subgroup identities. \textcolor{green!55!black}{\checkmark} & Attr & 1.79 \\
Race-consciousness is itself a cause of division and discrimination. \textcolor{green!55!black}{\checkmark} & Bel & 1.63 \\
Race is a superficial trait with no bearing on worth, ability, or character. \textcolor{green!55!black}{\checkmark} & Bel & 1.61 \\
\addlinespace
\multicolumn{3}{@{}l}{\footnotesize\itshape Intersectionality} \\
Collective liberation \textcolor{green!55!black}{\checkmark} & Val & 1.09 \\
Justice as the redress of structural inequality \textcolor{green!55!black}{\checkmark} & Val & 0.93 \\
Solidarity across marginalized groups \textcolor{green!55!black}{\checkmark} & Val & 0.92 \\
\bottomrule
\end{tabular}
\caption{Most important concepts per ideology in terms of $\eta$. Green checks indicate which agree with expectations.}
\label{tab:nodes}
\end{table}

We begin with Table~\ref{tab:ladder}, which shows results for a series of  models varying on a subset of model parameters. All models shown are run with no actor metadata, models are informed only by concepts extracted from tweets, pinned actors, and the logical conceptual constraints. Overall, Table~\ref{tab:ladder} shows that a model with one facet per ideology and two ideologies (IS and CB) with pinned actors best balances across the four evaluation metrics: all modeled ideologies are structurally coherent and have a non-zero share of actors, the model has the highest $\overline{AUC}$, and held-out predictive performance is competitive despite the model having the fewest parameters. 

Overall, the table reports evaluation statistics from three different sets of model parameterizations. The first models only two ideologies with pinned actors, IS and CB, but varies the number of facets per ideology. As noted, taken as a whole, the evaluation metrics suggest that a simple model with one facet per ideology performs well.  The second set of model parameterizations includes models with all three theorized ideologies, pinning 9 actors in total, three each to IS, CB, and WS. Consistent with expectations that white supremacist content is rare on Twitter \cite{siegelTrumpingHateTwitter2021}, we find no setting in which all three ideologies are structurally coherent and have a non-zero share of actors, suggesting neither model had a strong fit to the data. Finally, we consider the possibility that a third ``free'' ideological structure that we did not theorize in advance exists; we again find limited evidence of this.

Figure~\ref{fig:model_out_layers} presents the inferred ideological networks for the model with one facet for CB and IS (the top model in Table~\ref{tab:ladder}), and Table~\ref{tab:nodes} displays the top three nodes in each ideology from $\eta_k$. A few points are of interest. Table~\ref{tab:nodes} shows that the most important nodes for these ideologies exist \emph{at different layers of the network}. The top three nodes for intersectionality are all values, the top three for color-blindness are an attitude and two beliefs. Second, Table~\ref{tab:nodes} also reflects qualitatively what the $\overline{AUC}$ metric in Table~\ref{tab:ladder} shows quantitatively: our model is able to accurately assign these top concepts in ways that aligned with domain knowledge. Our model can thus recover concept to ideology associations by providing only domain knowledge about actors. Importantly, however, this is made easier by the closed-domain nature of our analysis; the model did not have to \emph{discover} these concepts, only assign them correctly to ideologies. 

Figure~\ref{fig:model_out_layers} shows two other points worth noting in the context of structure. First, while rationalizations were associated with concepts through their activations ($\eta_k$), there are almost no concept relationships involving rationalizations. This reflects the challenges noted above with sparsity of observation, but also the size of the dataset used for this working example.  Second, we observe that it is not only that concept activation rates differ across concept types for the ideologies, IS and CB are also distinct in their structural distribution. More specifically, while the most active concepts for IS are values, the value-to-value network is much denser for CB (0.32) than IS (0.048), whereas the belief-to-belief network is much denser for IS (0.32) than CB (0.11). This finding that the strength of a concept's association with an ideology therefore tells us something different from its structural position aligns with recent work challenging common assumptions of existing models of belief systems identified using survey data  \cite{fishmanChangeWeCan2022}.

Finally, while the above results do not leverage actor metadata, our model is capable of doing so. In the appendix, we provide one example of this, where we use regular expressions to identify themes embedded in user profile descriptions as signals of identity. We find that the model is capable of aligning these identity signals to ideologies in ways that are face-valid; for example, the model most strongly associates having MAGA-related keywords to CB, and BLM-related keywords to intersectionality. 

\section{Discussion and Conclusion}


\subsection{Contributions and Future Directions}\label{sec:insights}

\paragraph{Relation to dimensional models.} Our framework foregrounds the fact that ideology detection in online discourse involves simultaneous uncertainty over (1) whether a given ideology exists as a coherent system rather than an analyst-imposed category, (2) which concepts and relations constitute that ideology in a given socio-temporal context, and (3) whether particular discursive expressions reflect ideological commitment as opposed to other processes such as identity performance or framing. One objection to broadening ideology detection in this way is that it sets aside dimensional models that explain considerable variation in behavior and have a long history in our field \cite{laverExtractingPolicyPositions2003,barberaFollowYourIdeology2016}. We do not dispute the utility of such models: for researchers who need a highly generalizable measure of ideology as a dependent variable in the United States, the left/right model is difficult to improve upon \cite{jostPoliticalIdeologyIts2009}. Yet even in the hyper-partisan American context, a single dimension fails to explain the attitudes of a substantial share of the population \cite{kalmoeUsesAbusesIdeology2020}, and even two-dimensional models increasingly diverge from how major survey organizations such as Pew describe the electorate \cite{kileyRedVsBlue2026}. These models fare worse still globally \cite{maynardConvergenceDivergenceStudy2018}, and there are normative reasons to resist reducing politics to a low-dimensional structure \cite{wangSystemsFrameworkRemedying2021}.

\paragraph{Relation to belief-system models.}
Similar arguments for conceptual models of ideology are developed at length in a growing literature on quantitative models of belief systems, including work that bridges individual and socially shared beliefs \cite{dalegeNetworksBeliefsIntegrative2024}, represents concept hierarchies \cite{vlasceanuNetworkApproachInvestigate2024}, and links belief systems explicitly to left/right ideology \cite{warnckeResINNewMethod2026}. Notably, \citet{warnckeResINNewMethod2026} show that conceptual models offer a richer analytic space than the item-response models underlying most dimensional measures while still \emph{recovering their positionings}, showing that both present useful information and thus that conceptual and dimensional approaches are complements rather than rivals. 

Our framework builds on this work but differs in two respects. First, few existing models define ideologies over multiple distinct concepts. The closest such model is from \citet{vlasceanuNetworkApproachInvestigate2024}; we draw from this work but expand it by considering a richer model of what it means for an ideology to be coherent, incorporating structural and temporal factors. We also provide additional evidence that such multi-level network representations are useful in the study of online discourse: we find that different racial ideologies have their most important concepts at different conceptual levels (values for intersectionality, and attitudes/beliefs for color-blindness), and that the structural patterns differ between the ideologies (intersectionality has strong structure at the belief level, while color-blindness has stronger connections at the value level), suggesting that both the ``multi-level'' and the ``network'' nature of our model may be important to explore empirically.

\paragraph{Scaling the conceptual study of ideology.}
Still, perhaps the most practical contribution of the framework is that it allows the conceptual models of ideology discussed above, which focus primarily on survey data (where they use data at all) to scale to new data. Online discourse is abundant in ways that survey data are not, but extracting ideology from it involves a tension between \emph{description} and \emph{interpretation}. Descriptive measures, such as explicit issue stances \cite{caceres-wrightExplicitStanceDetection2024}, are more precise but also more sparsely expressed. Interpretive measures, such as approaches using natural language inference \cite{burnhamStanceDetectionPractical2025}, can be applied anywhere, but at the cost of noise. This is roughly the discursive analogue of the distinction between micro- and macro-level survey measures in \citet{berinskyMeasuringPublicOpinion2017}, who advocates measuring at an intermediate ``mezzo'' level. We instead propose modeling all levels jointly and using network structure to account for their dependencies, so that sparse but precise descriptive signals can constrain noisy interpretive ones, and vice versa.

\paragraph{Guiding the design of CSS studies.}
The framework is also useful before any model is fit. Treating ideology as a construct composed of multiple types of concepts and their relations makes explicit what a given design leaves out. A belief-network analysis, for example, may reveal dense structure without considering that this structure could arise from beliefs being anchored to a small set of underlying values. Conversely, a focus only on moral values captures those values but misses the concrete, falsifiable beliefs and justifications attached to them, which are often the most tractable targets for change. As with surveys \cite{berinskyMeasuringPublicOpinion2017}, work that focuses only on one level of abstraction is still useful and insightful, but should be clear about what such different levels of abstraction might make more or less salient. 

\paragraph{Implications for intervention.}
Finally, we hope the framework can help identify substructures within ideologies, such as highly central concepts or critical links, that are promising targets for intervention, particularly against harmful ideologies. As \citet{vlasceanuNetworkApproachInvestigate2024} argue, more successful interventions require mapping beliefs as elements of complex systems at both individual and collective levels. Recent experimental evidence also supports this: counterarguments targeting the beliefs people use to justify their attitudes produce larger and more durable attitude change than counterarguments targeting equally issue-relevant beliefs they did not invoke \cite{velezBeyond}. Knowing an ideology's structure, and which beliefs carry the most weight within it, is therefore not only descriptively interesting but practically consequential; network measures such as centrality offer one natural starting point. We recognize that the same tools could be used to target vulnerabilities in productive ideologies. As with many methods in CSS, however, we make our framework public in the expectation that its benefits to researchers and practitioners seeking to understand and improve online discourse outweigh the risk of misuse.

\subsection{Limitations}

These contributions aside, our work has significant limitations. These are most evident in our working example, where the most notable limitation is that any given result could be driven by extractor error. On the one hand, results for $\overline{AUC}$ suggest that the extractor at least functioned well enough to recover concept alignment. On the other, future work aiming to make concrete substantive claims would require validation via manual annotation. To this end, we provide in the appendix a potential way to use our approach  to inform this manual annotation process, building on ideas of active statistical inference presented by \citet{gligoricCanUnconfidentLLM2025}. This is because the model provides a theoretically and data-informed approach to identifying annotations that are incorrect. For example, our zero-shot extraction pipeline confidently identifies a tweet from a prominent advocate for intersectionality stating ``Race is not real. Anti-Black racism is real'' as expressing the colorblind belief that \emph{Race is a superficial trait with no bearing on worth, ability, or character}. Our statistical model, in contrast, attributes the odds of this actor expressing this belief as near-zero. Following the approach of \citet{gligoricCanUnconfidentLLM2025}, we can use these differences in scores between the statistical model and the extractor within an active inference setup to efficiently select posts for manual annotation for further fine-tuning, post-prediction inference \cite{egamiUsingImperfectSurrogates2023,angelopoulosPredictionPoweredInference2023}, or both.

There are also a number of less obvious limitations that would need to be addressed in a more comprehensive study. As noted, a more rigorous analysis would leverage newer methods, such as those from \citet{introneMeasuringBeliefDynamics2023} and \citet{burnhamStanceDetectionPractical2025}, that do not require an a priori defined list of ideological concepts. Doing so would likely involve mixed-methods approaches that make use of both large language models and qualitative analysis \cite{schroederLargeLanguageModels2025}. Further, such work would need to treat the boundaries between concept types as an empirical question. We rely on cases where a base of literature makes a clear-cut distinction between concept types (e.g. beliefs are falsifiable statements, whereas values are unfalsifiable but motivating), but real-world discourse often sits in a grey area between these well-defined boundaries. Learning where the boundaries fall in practice for a given dataset is an interesting area of future work.  A more robust analysis would also require better measures of uncertainty, such as from many runs of the model over partitions of the data, and a more robust exploration of the parameter space. Finally, our work could also be validated with data entirely external to the model but that should measure similar ideological patterns, such as the World Values Survey \cite{inglehart2000world}.

\section{Acknowledgements}

Generative AI was used as part of the methodology (in the zero-shot learning and the mixed methods approach to defining the codebook). More importantly, it was  used to create an inital draft of the write-up for the generative story and evaluation details for the working example, and to help write the code used to run the model.  KJ was supported by NSF IIS2145051.

\clearpage

\section{Paper Checklist}
\begin{enumerate}

\item For most authors...
\begin{enumerate}
    \item  Would answering this research question advance science without violating social contracts, such as violating privacy norms, perpetuating unfair profiling, exacerbating the socio-economic divide, or implying disrespect to societies or cultures?
    \answerYes{Yes, the aim, like many papers in CSS, is to focus on work that addresses these concerns by helping to shape our understanding of a social construct (ideological differences) and how to build a way to address the challenges it surfaces for society.}
  \item Do your main claims in the abstract and introduction accurately reflect the paper's contributions and scope?
    \answerYes{Yes.}
   \item Do you clarify how the proposed methodological approach is appropriate for the claims made? 
    \answerYes{Yes.}
   \item Do you clarify what are possible artifacts in the data used, given population-specific distributions?
    \answerYes{Yes.}
  \item Did you describe the limitations of your work?
    \answerYes{Yes.}
  \item Did you discuss any potential negative societal impacts of your work?
    \answerYes{Yes.}
      \item Did you discuss any potential misuse of your work?
    \answerYes{We discuss the potential that the framework could be used by bad actors to target vulnerabilities in productive ideologies, such as intersectionality. However, we emphasize that, as with many tools in CSS, we make our ideas and findings public with the expectation that the benefits to well-intentioned actors of understanding online discourse and potential ways of shifting it outweigh the drawbacks of fearing use of these tools by malign actors. }
    \item Did you describe steps taken to prevent or mitigate potential negative outcomes of the research, such as data and model documentation, data anonymization, responsible release, access control, and the reproducibility of findings?
    \answerYes{Yes.}
  \item Have you read the ethics review guidelines and ensured that your paper conforms to them?
      \answerYes{Yes.}
\end{enumerate}

\item Additionally, if your study involves hypotheses testing...
\begin{enumerate}
  \item Did you clearly state the assumptions underlying all theoretical results?
    \answerNA{N/A}
  \item Have you provided justifications for all theoretical results?
       \answerNA{N/A}
  \item Did you discuss competing hypotheses or theories that might challenge or complement your theoretical results?
     \answerNA{N/A}
  \item Have you considered alternative mechanisms or explanations that might account for the same outcomes observed in your study?
       \answerNA{N/A}
  \item Did you address potential biases or limitations in your theoretical framework?
      \answerNA{N/A}
  \item Have you related your theoretical results to the existing literature in social science?
      \answerNA{N/A}
  \item Did you discuss the implications of your theoretical results for policy, practice, or further research in the social science domain?
      \answerNA{N/A}
\end{enumerate}

\item Additionally, if you are including theoretical proofs...
\begin{enumerate}
  \item Did you state the full set of assumptions of all theoretical results?
      \answerNA{N/A}
	\item Did you include complete proofs of all theoretical results?
       \answerNA{N/A}
\end{enumerate}

\item Additionally, if you ran machine learning experiments...
\begin{enumerate}
  \item Did you include the code, data, and instructions needed to reproduce the main experimental results (either in the supplemental material or as a URL)?
      \answerYes{Yes, as an attachment and eventually, when author anonymity is no longer necessary, as a Github repository.}
  \item Did you specify all the training details (e.g., data splits, hyperparameters, how they were chosen)?
       \answerYes{Yes.}
     \item Did you report error bars (e.g., with respect to the random seed after running experiments multiple times)?
      \answerNo{We provide a working example, and thus note in the paper that a more robust analysis of this kind is out of scope for our primarily theoretical paper}
	\item Did you include the total amount of compute and the type of resources used (e.g., type of GPUs, internal cluster, or cloud provider)?
   \answerYes{Yes.}
     \item Do you justify how the proposed evaluation is sufficient and appropriate to the claims made? 
    \answerYes{Yes.}
     \item Do you discuss what is ``the cost'' of misclassification and fault (in)tolerance?
      \answerYes{Yes.}
  
\end{enumerate}

\item Additionally, if you are using existing assets (e.g., code, data, models) or curating/releasing new assets, \textbf{without compromising anonymity}...
\begin{enumerate}
  \item If your work uses existing assets, did you cite the creators?
     \answerYes{Yes.}
  \item Did you mention the license of the assets?
     \answerNA{N/A}
  \item Did you include any new assets in the supplemental material or as a URL?
   \answerYes{Yes.}
  \item Did you discuss whether and how consent was obtained from people whose data you're using/curating?
    \answerNo{We reference a full paper-length treatment of these important questions.}
  \item Did you discuss whether the data you are using/curating contains personally identifiable information or offensive content?
     \answerYes{Yes, we provide only de-identified data to address this concern.}
\item If you are curating or releasing new datasets, did you discuss how you intend to make your datasets FAIR (see \citet{fair})?
 \answerNA{N/A}
\item If you are curating or releasing new datasets, did you create a Datasheet for the Dataset (see \citet{gebru2021datasheets})? 
 \answerNA{N/A}
\end{enumerate}

\item Additionally, if you used crowdsourcing or conducted research with human subjects, \textbf{without compromising anonymity}...
\begin{enumerate}
  \item Did you include the full text of instructions given to participants and screenshots?
     \answerNA{N/A}
  \item Did you describe any potential participant risks, with mentions of Institutional Review Board (IRB) approvals?
 \answerNA{N/A}
  \item Did you include the estimated hourly wage paid to participants and the total amount spent on participant compensation?
     \answerNA{N/A}
   \item Did you discuss how data is stored, shared, and deidentified?
    \answerNA{N/A}
\end{enumerate}

\end{enumerate}

\appendix

\section{Appendix}

\subsection{Additional Definitions of Ideology}

\begin{table*}[t]
\footnotesize
\centering
\renewcommand{\arraystretch}{1.2}
\begin{tabularx}{\textwidth}{@{}p{2.5cm}p{2.3cm}X@{}}
\toprule
\textbf{Reference} & \textbf{Discipline} & \textbf{Quote} \\ 
\midrule

\cite{leadermaynardDangerousSpeechDangerous2016} & Ideology Studies &
A distinctive \textbf{system of} normative and/or purportedly factual \textbf{ideas}, typically \textbf{shared} by members of groups or societies, which \textbf{underpins individuals’ understanding} of their political world and \textbf{guides their political behaviour.} \\

\cite{jostPoliticalIdeologyIts2009} & Social Psychology & \textbf{A set of beliefs about the proper order of society and how it can be achieved} [or similarly] ``the \textbf{shared} framework of \textbf{mental models} that \textbf{groups} of individuals possess that \textbf{provide} both \textbf{an interpretation} of the environment \textbf{and a prescription} as to how that environment should be structured'' ...[that is] represented in memory as a kind of schema, i.e. a learned knowledge structure consisting of an interrelated network of beliefs, opinions, and values 
\\

\cite{oliverWhatGoodIdea2000} & Sociology &  The concept of ideology focuses on \textbf{ideas}, on \textbf{their systematic relations to each other}, \textbf{and on their implications} for social and political action based on value commitments...Wilson develops the very useful trichotomy of the structural elements of ideology which Snow and Benford adopted: diagnosis (how things got to be how they are), prognosis (which should be done and what the consequences will be), and rationale (who should do it and why)...an ideology links a \textbf{theory about society} with a \textbf{cluster of values} about what is right and wrong \textbf{as well as norms about what to do}.\\

\bottomrule
\end{tabularx}
\caption{Selected Definitions and Conceptualizations of Ideology}
\label{tab:ideology_defs}
\end{table*}

Table~\ref{tab:ideology_defs} presents a few representative definitions of ideology for those who are less familiar with this area of research.

\subsection{Other Potential Methods for Concept Extraction}
\begin{table*}[t]
\centering
\small
\setlength{\tabcolsep}{4pt}
\begin{tabular}{p{0.18\textwidth} p{0.12\textwidth} p{0.16\textwidth} p{0.46\textwidth}}
\hline
\textbf{Work} & \textbf{Concept} & \textbf{Task} & \textbf{Relevance to Our Framework} \\
\hline

\citet{introneMeasuringBeliefDynamics2023}
& Beliefs
& Belief identification
& Presents the Belief Landscape Framework, a mixed-methods approach that extracts belief statements using regular expressions and then uses semantic embeddings to cluster them into specific beliefs. The framework is interpretable and replicable, although it takes a broader definition of belief than we do here (and thus could potentially be used to extract attitudes as well). \\

\citet{borensteinInvestigatingHumanValues2025}
& Values
& Value classification
& Identifies expressions of the ten values defined by Schwartz's influential theory of values using social media data. This is a closed-domain approach that assumes a small, fixed set of values and constructs a supervised classifier in order to identify value expression. \\

\citet{huangValuesWildDiscovering2025}
& Values
& Value identification
& Proposes an open-domain approach to value identification in which values expressed in discourse are extracted verbatim and then clustered using a mixed-methods approach. Similar conceptually to the work of \citet{introneMeasuringBeliefDynamics2023}, but for values instead of beliefs.\\

\citet{burnhamStanceDetectionPractical2025}
& Attitudes
& Stance detection
& Uses natural language inference (NLI) to identify expressed issue stance. A text is paired with a hypothesis, and the NLI model estimates whether the text provides evidence for that claim. By definition (and as discussed in the main text), this method is inferential (rather than descriptive), and aims to tie stance detection directly to attitude expression through this inference process rather than only to expressed issue stance. \\

\citet{caceres-wrightExplicitStanceDetection2024}
& Attitudes
& Stance detection
& Focuses on extracting \emph{explicit} stance expressions, identifying cases where authors directly state a position on an issue. This contrasts with the inferential approach from \citet{burnhamStanceDetectionPractical2025}: explicit stance provides sparser but potentially stronger signals of expressed issue stance, but does not theoretically commit to any clear link between expressed issue stance and attitude, focusing only on whether or not a text expresses a particular stance. \\

\citet{barron-cedenoOverviewCheckThat20202020}
& Beliefs and rationalizations
& Claim and evidence retrieval
& Reviews techniques for automatic claim identification and verification. Claim retrieval seeks falsifiable statements, corresponding closely to our notion of beliefs, while evidence retrieval identifies evidence supporting or refuting claims, corresponding to our notion of rationalizations. These approaches could therefore provide components for open-domain extraction of beliefs and rationalizations. \\

\hline
\end{tabular}
\caption{Recent approaches to extracting beliefs, values, attitudes, and rationalizations from discourse.}
\label{tab:concept_extraction}
\end{table*}

Table~\ref{tab:concept_extraction} provides a sample of papers that describe methods that could be used to identify expressed ideological concepts in discourse.  In addition to providing a citation to the paper, the table notes what concept type(s) the methods presented in the paper are relevant to, the name of the NLP task(s) conducted, and a brief summary of the paper's methodology and its application to our framework.  Note that the table is far from exhaustive, and is only meant to present a sample of possible methods.

\subsection{Additional Potential Concept Types}

While our framework focuses on four concepts we find to be most often associated with ideologies, a number of other potential types of concepts could reasonably have been included. Narrower subtypes of the concepts in Table~\ref{tab:concept_defs}, like stereotypes (beliefs about individuals who hold particular social identities) and prejudices (attitudes toward such individuals) \cite{correllMeasuringPrejudiceStereotypes2010} are one example of a finer grain of detail our framework does not cover. Other potential concepts include a more generic model of perceptions, or what \citet{rayTheoryRacializedOrganizations2019} refers to as (socially shared) cognitive heuristics, which he sees as being an ideological component central to how racism emerges in organizations. The framework could also reasonably be extended to include (socially shared) emotions, which \citet{freeden2013emotions} argues can be ideological in the way that they are attached to political topics and/or entities, or more specifically in the way that they are a component of attitudes. In this sense, the concepts proposed here implicitly include emotion, in that we model item-responses for attitudes that could be expanded to, for example, different emotions surrounding affirmative action (e.g. anger), rather than simply support or opposition. But it would also be reasonable to include emotions as their own nodes in the network model. 

Other concepts that have been seen as more foundational to ideologies could also be represented, such as personality \cite{jostPoliticalIdeologyIts2009}. This also could include moral foundations from moral foundations theory (MFT). While many operationalize the underlying dimensions of MFT as values (and thus as being included already within our framework) \cite{jungVarietiesValuesMoral2025}, a more general focus on the notion of morality and its underlying connections to values and other ideological concepts may prove fruitful as well. Finally, as noted in the main text, it may also be interesting to further theorize not only the nodes and edges in the framework, but also meso-level constructs that likely involve multiple nodes and edges, such as conspiracy theories. Conspiracy theories are unique in that, much like ideologies, they are best theorized as networks of concepts that ``draw[s] its coherence from central beliefs such as the conviction that authorities and officials engage in massive deception of the public to achieve their malevolent goals'' \cite{woodDeadAliveBeliefs2012}.


\subsection{Data For Working Example}

\begin{table}[t]
\begin{center}\footnotesize
\begin{tabular}{@{}llr@{}}
\toprule
Ideology & Account & Units \\
\midrule
Color-blindness    & \texttt{PrisonPlanet}   & 1{,}070 \\
                   & \texttt{RealCandaceO}   &    447 \\
                   & \texttt{bluelivesmtr}   &    172 \\
\addlinespace
Intersectionality  & \texttt{deray}          & 3{,}900 \\
                   & \texttt{BreeNewsome}    & 2{,}386 \\
                   & \texttt{jemelehill}     &    270 \\
\bottomrule
\end{tabular}
\end{center}
    \caption{Actors pinned to IS and CB.}
    \label{tab:pinned}
\end{table}

We begin with a set of over 55M tweets sent between 2009-2024 that were retweeted by a panel of roughly 1.6 million Twitter accounts linked to U.S. voter registration records in 2018. Details on the methods used to link Twitter accounts to voters, their validity, and a summary of account behavior can be found in \citet{shugarsPandemicsProtestsPublics2021}. We then further subset our analysis 1) race-related retweets of tweets that contain a validated keyword list from \citet{shugarsPandemicsProtestsPublics2021}, and 2) specifically to retweets of accounts that had at least ten tweets retweeted at least once by the panel accounts, and were retweeted by at least 50 panel members. These accounts represent a set of actors that were influential in discourse about race on Twitter during this time period. 

Because we model tweet text, we retain only retweets that contain no URLs; six of the sampled accounts have no such tweets and are dropped, leaving 609 accounts. Two further accounts are dropped because the work here focuses only on the years from 2012--2022, where we have a stable stream of data, giving a final set of 607 actors and 62{,}045 coded tweets. Nine of these accounts serve as pinned actors, three per ideology: six are drawn from the list of known actors and three from the community samples. The reported analysis models colorblindness and intersectionality, so six pins are active in it; the three white-supremacy pins enter only in the appendix arm that models that component. Pinned actors for CB and IS are displayed in Table~\ref{tab:pinned}; actors pinned to WS are withheld because they are not clearly public actors. However, their usernames will be provided by the first author upon request.

\subsection{Additional Details on Concept Extraction}

In addition to extracting out concepts and providing a confidence score as outlined in the main section of the paper, we prompt the LLM to provide two additional fields. First, we ask the model to state whether the concept extracted is aligned with the author of the tweet stating their own thinking or if they are saying what they think someone else thinks. Second, we ask the model to determine whether the author holds the concept or rejects it. For attitudes, where both support and opposition for, e.g., housing reform are modeled as concepts, we prompt the model to select only one. Each concept extracted from a tweet can therefore fall into one of four categories: the author holds the concept, the author opposes/rejects the concept, the author reports that others hold it, or the author refutes it or attributes it to an opponent. In the work presented here, only the first (when the model infers that an author holds a concept) is modeled, all other observations are not given to the model. Of the 73,112 concept annotations in our corpus, 58,815 (80.4\%) record the author holding the concept; the remaining 14,297 are set aside but may be of interest for future work. 
 
Table~\ref{tab:codebook} presents the full set of concepts used in our working example. Note that many concepts are assigned to multiple ideologies; for example, White Supremacy (WS) and Intersectionality (IS) share beliefs surrounding the salience of race in public discourse. This is intentional, and reflects the fact that ideologies can be understood not only by a single concept, but by a set of concepts and their relations. As noted in the main text, to construct this table we draw on our prior work \cite{williamsBlackLivesMatter2024}, where we conduct a qualitative analysis using a grounded theory approach of a sample of tweets from the larger dataset from which our working example draws. This grounded theory approach surfaced a broad range of ideological concepts and tweets aligned with the three ideologies studied here. We then prompted Claude Fable 5.0 with the results from our grounded theory to generate a list of concepts that could be extracted using a zero-shot learning approach. We iterated this process and eventually manually finalized the final list of concepts in Table~\ref{tab:codebook}. As with the rest of the working example, we do not take this listing to be a complete or representative sampling of concepts representing these ideologies. However, it does cover the concepts that were most prominent from our grounded theory analysis.

The full prompt used for zero-shot labeling is given in Figure~\ref{fig:prompt}.  As noted in the main text, we use \texttt{openai/gpt-oss-20b} to perform the labeling. This model requires roughly 16GB of VRAM, and so our work can be run on a single standard GPU, although we run the model on a larger NVIDIA DGX B200 server that hosts several different models at our university. Because the model is run locally, there is no monetary cost to replicating the proposed approach beyond the cost of a GPU.

\subsection{Full details of the proposed model}

We define the set of concepts $\CC$ where each concept has a type defined by a type map $T:\CC\to\{V,A,B,R\}$. As mentioned in the main text, we make use of the item-response idea from \citet{warnckeResINNewMethod2026}, and thus model valences of attitudes (e.g., \emph{supports affirmative action} and \emph{opposes affirmative action} are distinct concept nodes). We use $u$ to represent a particular node throughout. With respect to relationships between concepts, we define \emph{allowed pairs} $\PP$ to be internal to a single concept type or between successive layers of the hierarchy:
\[
\begin{aligned}
\PP=\big\{(u,u'), \\[-2pt]
&\{T(u),T(u')\}\notin
\{\{V,B\},\{V,R\},\{A,R\}\}\big\}.
\end{aligned}
\]
As discussed in the main text and below, we also implement logical conceptual coherence by not allowing certain concepts that could not logically be in the same ideology (supporting and opposing affirmative action) to be paired together. 

The other inputs to the model are actor metadata $m_a$; tweet (discursive unit) timestamps $t(d)$, resolved to the yearly level here, a mapping from tweet author to tweet $a(d)$; the kernel $K_\delta$ (here, at the yearly level); any pinned actors $\Tcal^{\text{fix}}$; and the other hyperparameters for priors, introduced below as the parameters are introduced.
 
\subsubsection{Generative story}
 
\paragraph{Background.} For each node $u$, we first fix the value $m_u=\log(\text{corpus rate of }u)$, representing the base rate of the concept.
 
\paragraph{Ideologies and facets.} Each ideology $k=1,\dots,M$ is represented by $F_k$ facets. For each facet $f$ and node $u$, a sparse profile deviation is defined using the scale-mixture prior from the SAGE model \cite{eisensteinSparseAdditiveGenerative2011},
\[
  \tau^{(I)}_{k,f,u}\sim\text{Jeffreys},\qquad \eta_{k,f,u}\sim\mathcal{N}\big(0,\ \tau^{(I)}_{k,f,u}\big),
\]
When the value $\eta_{k,f,u}=0$, the facet takes no position on a given concept. In general, because of the sparsity assumption, an ideology is expected to hold this ``no position'' on most nodes. The faceting approach is intended to model the fact that online ideologues may try to focus on a subset of ideological content \cite{wanPoliticalPunditsMaintenance2026}. In practice, as we discuss in the main text, this faceting approach does not clearly improve performance, perhaps because racial ideology is a niche of broader political discourse already. Regardless, facets are connected into a single ideology because they share the ideology's actors and concept network $W_k$. 

For each pair of concepts $(u,u')\in\PP$, we assume $W_{k,uu'}\sim\text{Exp}(\lambda_{\text{sp}})$. Where pairs are not allowed (e.g. between rationalizations and values), i.e. where pairs between two concepts are not in $\PP$, we fix $W_{k,uu'}=0$. A shared association layer is drawn identically: $W^{(0)}_{uu'}\sim\text{Exp}(\lambda_{\text{sp}})$ on $\PP$. This $W^{(0)}$ is available to every actor and absorbs co-occurence patterns that every actor makes. This implements the distinction, drawn in the main text from \citet{vandijkIdeologyMultidisciplinaryApproach1998} as well as \citet{vlasceanuNetworkApproachInvestigate2024}, between ideology and culture. The background component $k=0$ has one facet with $\eta_{0,1}=0$ and $W_0=0$.
 
\paragraph{Actors.} Actors are defined as a soft membership over (ideology, facet) pairs including the background coordinate,
\[
\begin{aligned}
\theta_a &= \mathrm{softmax}\big(\beta_k^\top m_a+\gamma_{a,kf}\big),\\
\gamma_{a,\cdot} &\sim \mathcal{N}(0,\sigma_\gamma^2 I),\\
\beta_{k,j} &\sim \mathcal{N}(0,\sigma_\beta^2).
\end{aligned}
\]
For actors that are pinned, we fix their values of $\theta_a$ to a specific ideology using a point-mass on the relevant $k$. Identity metadata informs \emph{which ideology} an actor is associated with; the facet an actor works within is determined at the actor level. Below, $\vartheta_{a,k}=\sum_f\theta_{a,kf}$ denotes the ideology marginal; $\theta_{a,0}$ is a free parameter of the model that can, in practice, soak up some of the noise from the extraction process. Each actor also draws a breadth $\mu_a\sim\mathcal{N}(0,\sigma_\mu^2)$; this in principle allows for actors to cover a varying ``width'' of the ideological space.
 
\paragraph{Discursive Expression.} For each tweet $d$ sent by actor $a=a(d)$ at time $t=t(d)$ that expresses concept $u$, we assume that:
\[
\begin{aligned}
x_{d,u} &\sim \text{Poisson}\big(r_{a,u}\big),\\
\log r_{a,u}
  &= \underbrace{m_u}_{\text{background}}
   +\underbrace{\sum_{k,f}\theta_{a,kf}\eta_{k,f,u}}_{\text{ideology and facet}}
   +\underbrace{\mu_a}_{\text{breadth}}.
\end{aligned}
\]
Note that tweets $d$ are conditionally independent given rates. In particular, in this model, we aim to simplify inference by not modeling within-unit co-occurrence. Instead, we focus on temporally-chunked (yearly, here) concept co-occurrences at the actor level, as described next.
 
\subsubsection{Concept Pairs}
 
For each actor and concept pair $(u,u')\in\PP$, the kernel-weighted pair count is modeled as:
\[
\begin{aligned}
A^{(\delta)}_{a,uu'}
  &= \sum_{d,d'} K_\delta\big(t(d)-t(d')\big)\,
     x_{d,u}\,x_{d',u'},\\
A^{(\delta)}_{a,uu'}
  &\sim \text{Poisson}\!\Big(
     N^{(\delta)}_a\,\bar r_{a,u}\,\bar r_{a,u'}\,
     e^{\Omega_{a,uu'}}\Big),\\
\Omega_{a,uu'}
  &= W^{(0)}_{uu'}+\sum_k\vartheta_{a,k}W_{k,uu'},\\
N^{(\delta)}_a
  &= \sum_{d,d'}K_\delta\big(t(d)-t(d')\big).
\end{aligned}
\]
The expression $\bar r_{a,u}$ represents the actor's fitted rate for that concept. In general $K_\delta$ interpolates between an endpoint where $K=\1[d=d']$, in which case co-occurrence is modeled as occurring only when two concepts are expressed in the \emph{same tweet}, and $K\equiv1$, where co-occurrence is modeled at the actor level over all tweets the actor sends. While one could, in practice, identify a medium that best fits the data, here we simply assume a kernel width of one year, i.e. $K_\delta=\1[\text{year}(d)=\text{year}(d')]$, so pair counts and exposures are computed per (actor, year) block and summed. Finally, note that because we have a culturally shared association network across all actors, $W_k$ represents excess co-occurrence only for actors in a given ideology $k$, above and beyond the shared layer.
 
Finally, as noted above, this approach of creating a pseudo-liklihood on concept co-occurrance simplifies inference at the cost of complicating the conceptual logic of the model. More specifically, there is no joint distribution over $(x,A^{(\delta)})$ that can be easily specified. Instead, our pair model is an \emph{estimating equation}: the same dataset is scored through two different paths,  once through its margins (a true likelihood) and once through a derived pair statistic against a Poisson working model whose canonical parameter is $\Omega$. The drawback of this is that computing uncertainty on $W$ is challenging in this model; we leave this particular question to future work.
 
\subsubsection{Objective}

 Given the above, the objective to be minimized (a negative log-likelihood for the rate terms, the pair estimating function, and the sparsity penalties) is:
\[
\begin{aligned}
\mathcal{L}
={}&\sum_{d,u}\Big[r_{a(d),u}
-x_{d,u}\log r_{a(d),u}\Big]\\
&+\sum_{a,(u,u')\in\PP}
\Big[N^{(\delta)}_a\bar r_{a,u}\bar r_{a,u'}e^{\Omega_{a,uu'}}
-A^{(\delta)}_{a,uu'}\Omega_{a,uu'}\Big]\\
&+\lambda_{\mathrm{sp}}\sum_{(u,u')\in\PP}
\Big[W^{(0)}_{uu'}+\sum_kW_{k,uu'}\Big]
+\mathcal{L}_{\mathrm{prior}}.
\end{aligned}
\]

This function is minimized subject to $W\ge0$ and the constraints identified below, with
$\mathcal{L}_{\mathrm{prior}}$ collecting the SAGE priors on $\eta_{k,f}$ and Gaussian priors on $\beta$, $\gamma$, and $\mu$.

\subsubsection{Estimation}\label{app:estimation}

 We perform estimation through block-coordinate descent. Every block is convex or has a closed form given the others.
\begin{enumerate}\itemsep2pt
  \item \emph{Membership step.} Posterior over $\theta_a$ (ideology $\times$ facet, plus background) from the regression prior, the facet-logit prior, and the actor's aggregate loss; ideology fixed for pinned actors, facets always free.
  \item \emph{$\beta$-step.} Ridge multinomial regression of ideology margins on metadata. Pinned actors contribute no gradient to $\beta$.
  \item \emph{$\eta_{k,f}$- and $\mu$-steps.} Sparse log-link Poisson regressions with SAGE variance updates; Newton steps linear in the number of active cells.
  \item \emph{$W$-step.} A nonnegative $\ell_1$-penalized Poisson generalized linear model on allowed pairs, jointly over $W^{(0)}$ and $\{W_k\}$: kernel-weighted counts as observations, $N^{(\delta)}_a\bar r_{a,u}\bar r_{a,u'}$ as exposure, $(1,\vartheta_a)$ as design---an intercept (culture) and alignment covariates (ideologies), which is what makes the culture/ideology split an estimation property rather than a definition.
\end{enumerate}
We ensure identifiability by pinning actors to named ideologies, although as noted in the main text, we do experiment with free components as well.  

All models reported here are fit on a commodity laptop in under 10 minutes, with at most 30 iterations of the above inference loop. 
  
\subsubsection{Constraints}\label{app:constraints}
 
We constrain the above model via assumptions about logical conceptual coherence and the pinned actors discussed above.
 With respect to the former, we set the valence pairs of each attitude ($W_{k,uu'}=0$ between \emph{supports~$X$} and \emph{opposes~$X$}, for the shared layer and every ideology).
 
\subsection{Additional Details on Evaluation}\label{app:evaluation}
 
We propose four evaluation metrics, each of which we provide slightly more detail on here.
\subsubsection{Held-out fit}\label{app:heldout}
  Before fitting, and independently for each actor $a$, 10\% of that actor's $n_a$ units are drawn without replacement and withheld.  In order to calculate held-out fit, tweets are collapsed onto an (actor, year) grid, with each cell $g$ storing the number of units $n_g$ it contains and the summed extractor confidences $X_{gu}$ per node. Writing $\hat z_{gu}$ for the fitted log rate, the exact value we compute is
\[
  \mathrm{HO}\;=\;\sum_{g\in\text{test}}\sum_u\Big[\,n_g\,e^{\hat z_{gu}}
  \;-\;X_{gu}\,\hat z_{gu}\Big],
\]
This is the Poisson negative log-likelihood without the data-only factorial term, and lower is better. Note that this measure is sensitive to parameter counts, in that there is no penalty for models that have significantly more free parameters. Still, we largely find in our experiments that simpler models perform reasonably well. 
 
\subsubsection{Concept-mapping AUC}\label{app:auc}
 
Because the model is not ``told'' which concepts have been mapped to which ideology in our codebook, we can use these assignments to test the ability of the model to recover expected links between concepts and ideologies where those ideologies have pinned actors (and are thus identifiable).  Let
$G_k$ define the set of concepts we use to define ideology $k$ in Table~\ref{tab:codebook}. Writing $\eta^{\max}_{k,u}=\max_f\eta_{k,f,u}$ for the fitted profile and ranking all $K$ concepts by it, we report the rank-based (Mann--Whitney) area under the ROC curve,
\[
\begin{aligned}
\mathrm{AUC}_k
={}&\frac{1}{|G_k|\,|G_k^c|}
\sum_{u\in G_k}\sum_{u'\notin G_k}
\Big[\1\{\eta^{\max}_{k,u}>\eta^{\max}_{k,u'}\}\\
&\qquad\qquad
+\tfrac12\,\1\{\eta^{\max}_{k,u}=\eta^{\max}_{k,u'}\}\Big],
\end{aligned}
\]
Here, with $0.5$ chance and ties taking average ranks.  There are two reasons to select this approach for concept evaluation. First, because $22$ of the $112$ concepts are aligned with more than one ideology, we would like a measure that can handle multi-label assignment. Second, because we are modeling continuous values, we use a rank-based statistic rather than anything that assumes a categorical assignment. In order to average across all modeled ideologies, we report the macro average
$\overline{\mathrm{AUC}}=|\mathcal{K}|^{-1}\sum_{k\in\mathcal{K}}\mathrm{AUC}_k$ over the ideologies the codebook labels. 

\subsubsection{Structural coherence}\label{app:coherence}
 
For ideology $k$ with support
$S_k=\{u:\max_f\eta_{k,f,u}>0\}$, we compute structural coherence as 
\[
  C_k=\frac{\sum_{(u,u')\in\PP,\;u\in S_k,\;u'\in S_k}W_{k,uu'}}
           {\sum_{(u,u')\in\PP,\;u\in S_k\,\oplus\,u'\in S_k}W_{k,uu'}},
\]
Here, $\oplus$ is exclusive or, so the denominator focuses on cases with concept edges where one endpoint is inside the ideology and the other is outside of it. A value of $C_k>1$ says that the ideology's concepts relate more strongly to one another than they do to concepts outside the ideology.  As noted in the main text, the measure is tightly connected to the weighted conductance measure, $wc$, common in network analysis. Specifically, $C_k=\frac{1-wc}{2wc}$; we use our measure here instead of conductance simply because we feel it more naturally aligns with our definition of structural coherence.  As with $wc$, we can define a permutation null; here, the only difference is that our null sampling distribution is more conservative, controlling for the \emph{type} composition as well as node degree.
 
More specifically, we draw $1000$ samples to form a null distribution that can be used to test whether or not a given ideology's network $W_k$ is more or less structurally coherent than we would expect by chance for a network with a similar set of each concept type and degree distribution. Each of $1000$ null supports is drawn to match $S_k$ in size \emph{and} composition. All $K$ concepts are bucketed by a two-part stratification: concept type  and, within type, the quartile of weighted degree in $W_k$. A draw then takes from each bucket exactly as many concepts as $S_k$ contains, sampled without replacement, and $C_k$ is recomputed with that support substituted into the \emph{same} fitted $W_k$.
 
One final complication is that a null draw can result in an infinite value. With $E$ the number of exceedances (including infinite draws) among $N$ admissible draws, we therefore report the add-one permutation $p$-value $p=\frac{1+E}{1+N}$. 
 
\subsubsection{The actor-share check}\label{app:share}
 
The three quantities above can all improve while the model ceases to actually assign any actors to a given ideology. We therefore report, for each component, the percentage of actors whose ideology margin $\vartheta_{a,k}$ is largest on it.  This check is primarily focused on model degeneracy.
 
\subsection{Details on model linking profile descriptions to ideologies}
 
The main text briefly discusses a version of the model where we explicitly connect expressed identity to ideological commitments via the use of keywords embedded in user profiles. Here, we provide more detail on this analysis.

The first step of this was to construct a set of regular expressions that matched common themes we observed in the user bios. After identifying the themes, we used a generative model (Claude Opus 5.0) to help construct and validate regular expressions for each identified theme. Because we intended for this to be only a working example, we do not argue that the identified themes are complete, nor that the regular expressions used are robust to error. They are simply meant to serve as an example set of identity signifiers that could be used.  We then run a version of our model using the same parameterization as the one used to provide the results in Figure~\ref{fig:model_out_layers}, with the exception that instead of using a single intercept term in the regression model involving actor metadata $m_a$ (as in the main text), we include binary indicators for each actor signifying whether or not each regular expression we constructed for their bio was matched or not.  In total, we model eleven binary indicators, each of which is associated with one coefficient in $\beta$. The model is fit on all 62{,}045 units with no holdout.

\begin{table*}[t]
\centering
\small
\setlength{\tabcolsep}{5pt}
\renewcommand{\arraystretch}{1.15}

\begin{tabular}{@{}l
                >{\ttfamily\scriptsize\raggedright\arraybackslash}p{10.5cm}
                rrrr@{}}
\toprule
\textbf{Covariate} &
\textbf{Pattern} &
\textbf{$n$} &
\textbf{CB} &
\textbf{IS} &
\textbf{CB--IS} \\
\midrule

MAGA
&
\#maga\textbackslash b \,|\, \#kag \,|\, maga\textbackslash b \,|\,
trump2020 \,|\, americafirst \,|\, wwg1wga \,|\, trumptrain \,|\,
draintheswamp \,|\, patriot\textbackslash b \,|\,
\textbackslash bkag\textbackslash b \,|\, qanon \,|\, qarmy \,|\, potus45
&
61 & $1.021$ & $-0.841$ & $1.862$
\\

Conservative
&
\textbackslash bconservative\textbackslash b \,|\,
\textbackslash brepublican\textbackslash b \,|\,
\textbackslash bgop\textbackslash b \,|\,
\textbackslash blibertarian\textbackslash b \,|\,
\textbackslash bconstitutionalist\textbackslash b
&
33 & $0.340$ & $-0.097$ & $0.437$
\\

Faith
&
\textbackslash bgod\textbackslash b \,|\, christian \,|\,
\textbackslash bfaith\textbackslash b \,|\, jesus \,|\, blessed \,|\,
\textbackslash bchurch\textbackslash b \,|\, catholic \,|\, believer \,|\,
\textbackslash bpastor\textbackslash b \,|\, scripture
&
49 & $-0.304$ & $-0.537$ & $0.233$
\\

Guns
&
\#2a\textbackslash b \,|\, \textbackslash b2a\textbackslash b \,|\,
\textbackslash bnra\textbackslash b \,|\, second amendment \,|\,
gun rights \,|\, concealed
&
13 & $0.145$ & $-0.067$ & $0.212$
\\

Pro-life
&
prolife \,|\, pro-life \,|\, pro life
&
13 & $-0.233$ & $-0.297$ & $0.064$
\\

Liberal
&
\textbackslash bliberal\textbackslash b \,|\,
\textbackslash bprogressive\textbackslash b \,|\,
\textbackslash bdemocrat\textbackslash b \,|\,
\textbackslash bdemocratic\textbackslash b \,|\,
\textbackslash bleftist\textbackslash b \,|\,
\textbackslash bsocialist\textbackslash b \,|\,
\textbackslash bdsa\textbackslash b
&
24 & $0.022$ & $-0.007$ & $0.029$
\\

Media Account
&
journalist \,|\, reporter \,|\, correspondent \,|\, editor \,|\,
columnist \,|\, anchor \,|\, newsroom \,|\,
\textbackslash bpress\textbackslash b \,|\, contributor @ \,|\,
staff writer \,|\, bureau \,|\, @nytimes \,|\, @washingtonpost \,|\,
@cnn \,|\, @msnbc \,|\, @foxnews \,|\, @politico \,|\, @axios \,|\,
@reuters \,|\, @npr
&
85 & $0.126$ & $0.090$ & $0.036$
\\

\#Resist
&
\#resist\textbackslash b \,|\, \#theresistance \,|\, the resistance \,|\,
\textbackslash bresister\textbackslash b \,|\, \#impeach \,|\,
impeach45 \,|\, impeachtrump \,|\, bluewave \,|\,
votebluenomatterwho \,|\, \#fbr \,|\, nevertrump
&
23 & $-0.106$ & $-0.108$ & $0.002$
\\

Family
&
\textbackslash bmom\textbackslash b \,|\, \textbackslash bdad\textbackslash b \,|\,
father \,|\, mother \,|\, husband \,|\, \textbackslash bwife\textbackslash b \,|\,
grandma \,|\, grandpa \,|\, nana \,|\, married \,|\, daughter \,|\, parent
&
68 & $0.306$ & $0.358$ & $-0.052$
\\

\#BLM
&
\textbackslash bblm\textbackslash b \,|\, black lives \,|\,
blacklivesmatter
&
23 & $-0.552$ & $0.086$ & $-0.638$
\\

Pronouns
&
she/her \,|\, he/him \,|\, they/them \,|\, she/they \,|\,
he/they \,|\, \textbackslash(pronouns
&
35 & $-0.811$ & $-0.078$ & $-0.733$
\\

\bottomrule
\end{tabular}

\caption{Details for the model linking user bios to ideological affiliations
using regular expressions applied to user bios. Each row shows a covariate
(theme), the regular expression used to identify the theme, the number of
matching user bios $n$ (bios may match multiple themes), and the coefficients
for colorblindness (CB) and intersectionality (IS). The final column reports
the difference $\mathrm{CB}-\mathrm{IS}$. Rows are sorted by this difference.}

\label{tab:kwbeta}
\end{table*}

Table~\ref{tab:kwbeta} shows the full regular expressions used for each of the eleven identifying themes we considered for user bios, as well as the coefficient linking that theme to colorblindness and intersectionality, and the difference between these two coefficients. Because each ideology is linked to these indicators through a separate regression model relative to a corpus baseline, the best way to interpret them is to look at the final column of the table, which shows the difference in the impact each theme has on association to each ideology. The table is also sorted by this value in order from indicators most associated with colorblindness to those most associated with intersectionality.  Notably, none of the six pinned actors in Table~\ref{tab:pinned} matches any of the regular expressions used, so the learned coefficients are driven only by what is learned about actors for whom the model has no initial indications about their ideological leanings from the model setup.  

Table~\ref{tab:kwbeta} primarily shows face validity of our work: the themes in user bios most closely tied to a colorblind ideology are those associated with the MAGA and conservative movements, whereas users who explicitly state their pronouns in their bios or an explicit affiliation with the intersectional Black Lives Matter movement (BLM) are most likely to express concepts relevant to an intersectional identity. All other coefficients are substantially smaller or near zero. Of course, in a more robust analysis, we would iterate over the model in order to provide measures of uncertainty, and/or, as described above, move to an approach where a true posterior could be computed in order to give credible intervals.

 \subsection{Using the model to select annotations}\label{app:annotation}

As emphasized in the main text, any result above could be driven by extractor error, and validating substantive claims
would require manual annotation. As also noted in the main text, the model provides a way to spend that annotation budget well. Confidence-driven inference \citep{gligoricCanUnconfidentLLM2025} selects instances for human
annotation using the LLM's own confidence, then combines human and LLM labels into estimates with valid confidence intervals, reducing the required number of human annotations. Our model supplies a second and independent signal for that selection: the fitted rate $r_{a,u}$ is the model's expectation that \emph{this actor} expresses \emph{this concept}, so an annotation that is confident under the extractor but improbable
under the model is a natural candidate for review. The example in the main text is of this kind, where the extractor confidently reads a colorblind belief in a tweet by a prominent intersectional advocate, while the model puts that actor's rate on the concept near zero. Scoring every annotation by the disagreement between extractor confidence and fitted rate
gives a ranked list to annotate.

Corrected labels could then be used in two ways. One version would retain the model proposed in this paper, correcting the extractor's output before fitting, or reporting corrected estimates of the quantities that depend most directly on rates. The more thorough option avoids the coupled estimation entirely: with reliable labels on a sample, the expression
and co-occurrence layers could be estimated in sequence rather than jointly, with the
annotated subset calibrating the extraction step before any ideological structure is fit.
We leave both to future work.

\onecolumn
\begin{longtable}{@{}llp{0.78\textwidth}@{}}
\toprule
ID & Ideology & Concept \\
\midrule
\endfirsthead
\caption[]{\emph{(continued)}}\\
\toprule
ID & Ideology & Concept \\
\midrule
\endhead
\midrule
\multicolumn{3}{r@{}}{\footnotesize\emph{continued on the next page}}\\
\endfoot
\bottomrule
\endlastfoot
\multicolumn{3}{@{}l}{\textit{Values} (13 concepts)}\\[1pt]
V1 & CB & Equality (the same rules and treatment for everyone) [Abstract Liberalism] \\
V2 & IS & Equity (differential treatment where needed to produce fair outcomes) \\
V3 & CB & Individualism (persons are to be judged as individuals, not as group members) [Abstract Liberalism] \\
V4 & CB & Personal responsibility (individuals are accountable for their own outcomes) [Abstract Liberalism] \\
V5 & CB & Meritocracy (rewards should track effort and ability) [Abstract Liberalism] \\
V6 & CB, WS & Freedom of choice [Naturalization] \\
V7 & CB, WS & Freedom of association [Naturalization] \\
V8 & WS & Racial hierarchy as natural order \\
V9 & WS & Racial purity \\
V10 & WS & Racial in-group loyalty and preservation \\
V11 & IS & Justice as the redress of structural inequality \\
V12 & IS & Solidarity across marginalized groups \\
V13 & IS & Collective liberation \\
\addlinespace[3pt]
\multicolumn{3}{@{}l}{\textit{Attitudes} (28 concepts)}\\[1pt]
A1 & CB & Disapproval of talking about race in schools, workplaces, and public life. [Race salience] \\
A2 & CB & Disapproval of people who identify strongly with their racial group. [Race salience] \\
A3 & WS, IS & Positive evaluation of explicit racial identification and racial group consciousness. [Race salience] \\
A4 & WS & Positive evaluation of white racial pride and white identity organizing. [Race salience] \\
A5 & CB, WS & Opposition to affirmative action. [Abstract Liberalism] \\
A6 & CB, WS & Opposition to reparations. [Abstract Liberalism] \\
A7 & CB, WS & Opposition to race-conscious policy generally. [Abstract Liberalism] \\
A8 & IS & Support for affirmative action. \\
A9 & IS & Support for reparations. \\
A10 & IS & Support for race-targeted public investment. \\
A11 & CB, WS & Skepticism toward DEI programs. [White Victimization] \\
A12 & CB, WS & Disapproval of 'identity politics.' \\
A13 & CB & Approval of class-based or need-based policies as legitimate substitutes for race-based ones. \\
A14 & CB & Positive evaluation of a shared national identity over subgroup identities. \\
A15 & WS & Support for the creation of a white ethnostate. \\
A16 & WS & Opposition to intermarriage across races. \\
A17 & WS & Opposition to civil rights law and anti-discrimination enforcement. \\
A18 & WS & Support for open hostility towards Black people. \\
A19 & WS & Contempt for whites who support racial equality ('race traitors'). \\
A20 & WS & Positive evaluation of historical figures associated with white rule. \\
A21 & WS, IS & Distrust of mainstream media, academia, and government. \\
A22 & IS & Support for structural reform of policing. \\
A23 & IS & Support for structural reform of housing. \\
A24 & IS & Support for structural reform of health systems. \\
A25 & IS & Positive evaluation of coalition politics linking racial, gender, economic, and disability justice movements. \\
A26 & IS & Skepticism toward colorblind or 'universal' policies as likely to reproduce existing hierarchies. \\
A27 & IS & Positive evaluation of representation of marginalized people in institutions. \\
A28 & IS & Positive evaluation of participatory decision-making. \\
\addlinespace[3pt]
\multicolumn{3}{@{}l}{\textit{Beliefs} (35 concepts)}\\[1pt]
B1 & CB & Race is a superficial trait with no bearing on worth, ability, or character. \\
B2 & WS, IS & Race is a central category for understanding how society works and how people fare in it. [Race salience] \\
B3 & WS & Race is meaningful because it is a biological reality. [Race salience] \\
B4 & IS & Race is meaningful because, although socially constructed, it structures institutions and life chances. [Race salience] \\
B5 & CB & Race should be de-emphasized in public discourse, policy, and personal identity. [Race salience] \\
B6 & WS, IS & Race should be made explicit and foregrounded in public discourse, policy, and identity. [Race salience] \\
B7 & CB & Race-consciousness is itself a cause of division and discrimination. [Race salience] \\
B8 & WS, IS & Ignoring race allows existing racial arrangements to persist unchallenged. [Race salience] \\
B9 & WS & Races are real, biologically fixed categories. \\
B10 & WS & Races can be ranked from superior to inferior. \\
B11 & WS & White people are biologically and intellectually superior to other races. [White Superiority] \\
B12 & WS & White people are culturally superior to other races. [White Superiority] \\
B13 & WS & Because they are superior, white people are entitled to dominate socially and politically. [White Superiority] \\
B14 & CB & Racism is essentially individual prejudice held by a small number of bad actors, not a feature of institutions. [Abstract Liberalism] \\
B15 & IS & Racism is embedded in the structure of American institutions (law, housing, education, health care, criminal justice). [Structural racism] \\
B16 & IS & Institutional racism operates regardless of the intent of the individuals within institutions. [Structural racism] \\
B17 & CB & Structural or institutional racism may exist in some residual form, but its effect on outcomes today is small. [Minimization] \\
B18 & CB, WS & Persistent racial disparities are mostly explained by differences in cultural values, work ethic, and family structure across groups. [Cultural Racism] \\
B19 & WS & Persistent racial disparities reflect innate differences in capacity. \\
B20 & IS & Persistent racial disparities are the product of historical policies and present structural conditions. \\
B21 & CB, WS & Residential and school segregation reflect people's natural preference to live among those like themselves. [Naturalization] \\
B22 & CB, WS & Discrimination against white people is now as common as, or more common than, discrimination against Black people. [White Victimization] \\
B23 & CB, WS & White people are under attack demographically, culturally, and politically. [White Victimization] \\
B24 & CB, WS & Race-conscious policies (affirmative action, race-based scholarships, diversity quotas) are themselves a form of racial discrimination. \\
B25 & IS & Social identities (race, gender, class, sexuality, disability, and so on) are multiple and intersecting, not independent or additive. [Multiple, intersecting identities] \\
B26 & IS & A person's social position is the product of how their identities combine, not the sum of each considered separately. [Multiple, intersecting identities] \\
B27 & IS & Intersecting identities include dominant as well as marginalized ones. [Identities need not be marginalized] \\
B28 & IS & Everyone occupies a position in overlapping systems of advantage and disadvantage. [Identities need not be marginalized] \\
B29 & IS & Systems of oppression (racism, sexism, classism, heterosexism, ableism) are interlocking and mutually reinforcing. \\
B30 & IS & Those at the intersections of multiple marginalizations experience harms that are invisible to analyses using any single category. \\
B31 & IS & Change that begins from the most marginalized improves conditions for everyone. [Margins as focal point] \\
B32 & WS & Civilization and its major achievements are the products of white peoples. \\
B33 & WS & Racial mixing degrades the white race. \\
B34 & WS & Demographic change threatens the survival of the white race. \\
B35 & WS & Institutions that promote racial equality are captured by hostile elites or out-groups. \\
\addlinespace[3pt]
\multicolumn{3}{@{}l}{\textit{Rationalizations} (36 concepts)}\\[1pt]
R1 & CB & The Civil Rights Act and the Fourteenth Amendment already make racial discrimination illegal, so anyone who is actually discriminated against can sue (and similar). [Abstract Liberalism] \\
R2 & CB & Jim Crow ended sixty years ago and nobody alive today owned slaves, so present-day inequality can't be blamed on racism (and similar). [Minimization] \\
R3 & CB & The country elected a Black president twice, so racism can't be holding Black people back (and similar). [Minimization] \\
R4 & CB & There are a lot of rich Black people (Oprah, LeBron James, Black CEOs), how could that happen if there was racism? (and similar) [Minimization] \\
R5 & CB & Surveys show almost no white people say they would object to a Black neighbor or a Black candidate, so prejudice has essentially disappeared (and similar). [Minimization] \\
R6 & CB, WS & Asian Americans immigrants out-earn white Americans, which shows that discrimination can't explain the Black-white gap, culture does (and similar). [Cultural Racism] \\
R7 & CB, WS & The high rate of single-parent households among Black families explains gaps in income, education, and incarceration (and similar). [Cultural Racism] \\
R8 & CB & Anyone who finishes high school, gets a job, and waits to have children until marriage will not be poor, regardless of race (and similar). [Cultural Racism] \\
R9 & CB, WS & Black people choose to live in Black neighborhoods, attend Black churches, and go to HBCUs, everyone prefers their own kind, so segregation isn't imposed (and similar). [Naturalization] \\
R10 & CB, WS & Universities admit Black applicants with lower test scores than white and Asian applicants they reject, which is discrimination against whites and Asians (and similar). [White Victimization] \\
R11 & CB, WS & A white man can't get hired or promoted anymore because companies have diversity targets (and similar). [White Victimization] \\
R12 & CB & Telling young people the system is rigged against them discourages them from trying, students told they are victims do worse (and similar). \\
R13 & CB & Martin Luther King Jr. said people should be judged by the content of their character, not the color of their skin, race-conscious policy betrays that vision. \\
R14 & CB & Kids don't notice race until adults teach them to, when schools and workplaces stopped talking about race, people just saw each other as classmates and coworkers (and similar). [Race salience] \\
R15 & CB & Race relations got worse after diversity training and race-focused curricula were introduced, which shows that talking about race creates the resentment it claims to cure (and similar). [Race salience] \\
R16 & WS & Every other group is allowed to organize openly as a race (Black caucuses, La Raza, Jewish organizations), so whites must do the same or be the only group without a voice (and similar). [Race salience] \\
R17 & WS & Colorblindness is a trick: whites are told to ignore race while every other group is organizing on it, which guarantees whites lose (and similar). [Race salience] \\
R18 & IS & After California banned race-conscious admissions (Prop 209), Black and Latino enrollment at Berkeley and UCLA fell sharply, showing that ignoring race locks in existing inequality (and similar). [Race salience] \\
R19 & IS & 'Race-neutral' policies such as legacy admissions, credit scores, or school funding by property tax reproduce racial gaps precisely because they don't account for race (and similar). [Race salience] \\
R20 & WS & Published IQ test averages differ by race (e.g., claimed 15-point Black-white gap), which proves innate intellectual differences (and similar). [White Superiority] \\
R21 & WS & Europeans produced the scientific revolution, the printing press, and modern medicine while sub-Saharan Africa built no comparable civilization, which proves white cultural superiority (and similar). [White Superiority] \\
R22 & WS & Black Americans are 13\% of the population but commit roughly half of homicides, which proves innate differences in behavior (and similar). \\
R23 & WS & Countries and cities that became majority non-white (e.g., Detroit, Zimbabwe, South Africa) declined afterward, which shows white people are needed to run a functioning society (and similar). \\
R24 & WS & Every animal species sticks to its own kind and defends its territory, so racial loyalty and separation are natural (and similar). \\
R25 & WS & The Census projects that whites will be a minority by the 2040s, and this was engineered through immigration policy by hostile elites (and similar). \\
R26 & WS & Mixed-race children have worse identity, health, or social outcomes, which shows racial mixing is harmful (and similar). \\
R27 & WS & The 1790 Naturalization Act limited citizenship to 'free white persons,' so the founders intended a white nation (and similar). \\
R28 & WS & Scripture identifies white Europeans as God's chosen people or as the true Israelites (and similar). \\
R29 & IS & Black women die in childbirth at roughly three times the rate of white women even when income and education are held constant, a harm visible neither in 'race' nor 'gender' data alone (and similar). [Multiple, intersecting identities] \\
R30 & IS & The Black-white wealth gap today maps directly onto 1930s redlining maps and onto exclusion from the GI Bill and FHA loans, showing that the gap was produced by policy (and similar). [Structural racism] \\
R31 & IS & The 100-to-1 crack-versus-powder-cocaine sentencing disparity produced racially skewed incarceration without any individual judge needing to be prejudiced (and similar). [Structural racism] \\
R32 & IS & Audit studies show white applicants with a criminal record receive callbacks at the same rate as Black applicants with no record, so discrimination persists even as stated attitudes improve (and similar). [Structural racism] \\
R33 & IS & A wealthy Black man is still more likely to be stopped by police than a poor white man, a white working-class woman is advantaged by race and disadvantaged by class and gender, so single identities cannot capture anyone's position (and similar). [Identities need not be marginalized] \\
R34 & IS & White women have been the largest beneficiaries of affirmative action, while the feminist movement largely ignored Black and working-class women's demands, showing that reforms aimed at a group's most advantaged members leave the rest behind (and similar). [Margins as focal point] \\
R35 & IS & Curb cuts built for wheelchair users ended up helping parents with strollers and delivery workers, designing for the most marginalized benefits everyone (and similar). [Margins as focal point] \\
R36 & IS & Domestic workers and enslaved people understood how the household and the plantation functioned better than their employers did, those who bear a system's burden see it most clearly (and similar). \\
\caption{A full list of the 112 concepts (values, beliefs, attitudes, rationalizations) used in our worked example in the paper. \emph{Ideology} lists the ideologies the concept is assigned to based on domain expertise. }
\label{tab:codebook}
\end{longtable}

\begin{figure*}[t]
\centering
\scriptsize

\fbox{%
\begin{minipage}{0.97\textwidth}
\textbf{Extraction prompt.}
Below is a codebook of ideological elements, grouped into four layers:
Values, Beliefs, Attitudes, and Rationalizations.

\texttt{\{\{codebook\}\}}

Here is a text (a tweet):

\texttt{"""\{\{tweet\}\}"""}

Which codebook elements does this text engage?

Code an element only if the text genuinely engages it. Do \emph{not} code
an element merely because the text is on the same topic, or because a word
from the element appears in it. Most texts engage a handful at most, and an
empty list is a fine answer.

For each element you code, answer two separate questions.

\textbf{WHOSE thinking is being stated?}
\begin{itemize}
\item \texttt{author}: the author is stating their own thinking.
\item \texttt{other}: the author is stating someone else's thinking---quoting,
describing, or summarising what another person, group, or movement says,
believes, or does.
\end{itemize}

\textbf{WITH WHAT STANCE is that position put?}
\begin{itemize}
\item \texttt{support}: the position is asserted or supported---put forward
as true, held as a value, endorsed, or backed in what it calls for.
\item \texttt{reject}: the position is rejected or refuted---denied, argued
against, mocked, or opposed in what it calls for.
\end{itemize}

Where the codebook carries a separate element for the position the author
actually holds (e.g., both ``Opposition to affirmative action'' and
``Support for affirmative action''), code that element with \texttt{support}
rather than its opposite with \texttt{reject}. Use \texttt{reject} only when
no element matches the position held.

The two fields are independent. An author can report someone else's position
approvingly (\texttt{other + support}) or attribute it to an opponent in order
to attack them (\texttt{other + reject}); they can hold a position
(\texttt{author + support}) or oppose it (\texttt{author + reject}).
Only \texttt{author + support} means the author holds the element. The other
combinations record that the element is present without the author standing
behind it.

\vspace{2pt}
\hrule
\vspace{2pt}

\textbf{Worked examples.}

\texttt{``The MAGABomber is a hardcore white supremacist who called Black
people slurs.''}

$\rightarrow$ racial superiority: \texttt{other + support}. Someone else's
position is reported; the author asserts none of it.

\texttt{``Biden is the only candidate with ties to the KKK and white
supremacy.''}

$\rightarrow$ racial superiority: \texttt{other + reject}. The position is
attributed to an opponent as a charge against them.

\texttt{``If you can read this then you're racist, because English was invented
by white devils. Checkmate.''}

$\rightarrow$ racial superiority: \texttt{other + reject}. The author mocks a
position attributed to others.

\texttt{``Affirmative action is unfair and should be scrapped.''}

$\rightarrow$ opposition to affirmative action: \texttt{author + support}.
When the codebook contains the element matching the author's position, code
that element rather than its opposite.

\texttt{``People keep saying Jim Crow was ages ago so racism is over.
Nonsense.''}

$\rightarrow$ Jim Crow being too long ago to explain inequality:
\texttt{author + reject}. Here the codebook has no separate element matching
the author's position, so \texttt{reject} records the opposition.

\vspace{2pt}
\hrule
\vspace{2pt}

\textbf{Output.}
Return only a JSON object with key \texttt{elements}, an array of objects
with keys \texttt{id}, \texttt{confidence}, \texttt{source}, and
\texttt{stance}:

\texttt{[{"id":"V1","confidence":0.9,"source":"author","stance":"support"},}

\texttt{\phantom{[}{"id":"B11","confidence":0.7,"source":"other","stance":"reject"}]}

\end{minipage}%
}

\caption{Prompt used to extract ideological elements from each text. The
codebook contains 112 concepts across four layers; \texttt{\{\{codebook\}\}}
is rendered as one concept per line, while \texttt{\{\{tweet\}\}} denotes the
unit being coded. The prompt separates the source of a position
(\texttt{author}/\texttt{other}) from its stance (\texttt{support}/\texttt{reject});
only \texttt{author + support} indicates that the author holds the coded
element.}
\label{fig:prompt}
\end{figure*}

\end{document}